\documentclass[preprint]{elsarticle}
\usepackage{lineno,hyperref}
\modulolinenumbers[5]

\usepackage{tabulary}
\usepackage{booktabs}
\usepackage{multirow}

\usepackage{graphicx}
\usepackage{dcolumn}
\usepackage{bm}
\usepackage{epstopdf}
\usepackage{mathtools}
\usepackage{natbib}
\usepackage{captcont}
\usepackage{xcolor}
\usepackage{ulem}
\usepackage{supertabular}
\usepackage{subfigure}

\journal{Astroparticle Physics}

\begin{document}

\begin{frontmatter}

\title{The on-orbit calibration of DArk Matter Particle Explorer}


\author[INFN-Perugia]{G. Ambrosi}

\author[USTC-Keylab]{Q. An}

\author[GU]{R. Asfandiyarov}

\author[GU]{P. Azzarello}

\author[U-Lecce,INFN-Lecce]{P. Bernardini}


\author[PMO]{M. S. Cai}

\author[INFN-Bari]{M. Caragiulo}

\author[PMO,USTC-Astron,UCAS]{J. Chang}

\author[PMO,USTC-Astron]{D. Y. Chen}

\author[USTC-Keylab]{H. F. Chen}

% ...

\address[PMO]{Key Laboratory of Dark Matter and Space Astronomy, Purple Mountain Observatory, Chinese Academy of Sciences, Nanjing 210034, China}
\address[INFN-Perugia]{Istituto Nazionale di Fisica Nucleare Sezione di Perugia, I-06123 Perugia, Italy}
\address[USTC-Keylab]{State Key Laboratory of Particle Detection and Electronics, University of Science and Technology of China, Hefei 230026, China}
\address[GU]{Department of Nuclear and Particle Physics, University of Geneva, CH-1211, Switzerland}
\address[U-Lecce]{Universit\`a del Salento - Dipartimento di Matematica e Fisica "E. De Giorgi", I-73100, Lecce, Italy}
\address[INFN-Lecce]{Istituto Nazionale di Fisica Nucleare (INFN) - Sezione di Lecce, I-73100, Lecce, Italy}
\address[INFN-Bari]{Istituto Nazionale di Fisica Nucleare Sezione di Bari, I-70125, Bari, Italy}
\address[USTC-Astron]{School of Astronomy and Space Science, University of Science and Technology of China, Hefei, Anhui 230026, China}
\address[UCAS]{University of Chinese Academy of Sciences, Yuquan Road 19, Beijing 100049, China}
\address[IMP]{Institute of Modern Physics, Chinese Academy of Sciences, Nanchang Road 509, Lanzhou 730000, China}
\address[NSSC]{National Space Science Center, Chinese Academy of Sciences, Nanertiao 1, Zhongguancun, Haidian district, Beijing 100190, China}
\address[GSSI]{Gran Sasso Science Institute (GSSI), I-67100, L'Aquila, Italy}
\address[IHEP]{Institute of High Energy Physics, Chinese Academy of Sciences, YuquanLu 19B, Beijing 100049, China}
\address[U-Bari]{Dipartimento di Fisica "M.Merlin" dell'Univerisity e del Politecnico di Bari, I-70126, Bari, Italy}
\address[U-Perugia]{Dipartimento di Fisica e Geologia, Universit\`a degli Studi di Perugia, I-06123 Perugia, Italy}
\address[UH]{Department of Physics, The University of Hongkong, Pokfulam Road, Hong Kong}

\begin{abstract}

The DArk Matter Particle Explorer (DAMPE), a satellite-based cosmic ray and gamma-ray detector, was launched on December 17, 2015, and began its on-orbit operation on December 24, 2015. In this work we document the on-orbit calibration procedures used by DAMPE and report the calibration results of the Plastic Scintillator strip Detector (PSD), the Silicon-Tungsten tracKer-converter (STK), the BGO imaging calorimeter (BGO), and the Neutron Detector (NUD). The results are obtained using Galactic cosmic rays, bright known GeV gamma-ray sources, and charge injection into the front-end electronics of each sub-detector. The determination of the boundary of the South Atlantic Anomaly (SAA), the measurement of the live time, and the alignments of the detectors are also introduced. {The calibration results demonstrate the stability of the detectors in almost two years of the on-orbit operation.}

\end{abstract}


\end{frontmatter}


\section{introduction}

DAMPE is a satellite-based, general-purpose, high-energy particle detector that can measure cosmic ray electrons and gamma-rays in the energy range from $\sim 5$ GeV to $\sim 10$ TeV and cosmic ray protons and heavier nuclei from tens of GeV to hundreds of TeV. 
With an unprecedentedly high energy resolution, a strong electron/hardron separation power, and a reasonably large acceptance, DAMPE is expected to considerably advance our understanding on the cosmic ray electron spectrum up to $\sim 10$ TeV and also play an important role in the gamma-ray line searches \cite{Chang2009,WuJ2011,Chang2014,Chang2017}.

The DAMPE instrument, from top to bottom, consists of a Plastic Scintillator strip Detector (PSD), a Silicon-Tungsten tracKer-converter (STK), a BGO imaging calorimeter (BGO), and a NeUtron Detector (NUD) \cite{Chang2014,Chang2017}. The PSD is composed of 82 organic plastic scintillators to form a double-layer $x$-$y$ detection array that covers an active area of $82~{\rm cm} \times 82~{\rm cm}$ \cite{YuYH2017}. The PSD measures the charge of incident cosmic rays and provides charged-particle background rejection for the gamma-ray observation \cite{YuYH2017}. The STK is composed of 12 silicon detector layers with coordinate readout in alternative orthogonal arrangement with a total area of about $7~{\rm m^2}$. Three thin tungsten layers are inserted to promote the photon conversion. 
The STK measures the charges and the trajectories of charged particles, and allows to reconstruct the directions of incident photons that have been converted into $e^{\pm}$ pairs in the STK material \cite{Azzarello:2016trx,Wu2015}. The BGO detector, composed of about 32 radiation lengths of BGO scintillation crystals arranged hodoscopically in 14 layers, measures the energy and distinguishes between electromagnetic and hadronic showers \cite{ZhangZY2016,ZhangZY2015,DongJN2017}. The NUD can further improve the lepton/hardron identification capability \cite{HeM2016}. DAMPE has in total 75908 detector readout channels. The beam-test results as well as the current DAMPE performance are described in \cite{Chang2017,ZhangZY2016,YueC2017,DAMPE2017,XuZL2017a,DuanKK2018,MaPX2018,DongTK2018,LiuY2017,ZhangYP2017,WuX2017,Vitillo2017,FengCQ2017,ZangJJ2017,LeiSJ2017,LiangYF2017,YuanQ2017,Fernanda2017,YueC2017b,Gallo2017}.


The on-orbit calibration of each sub-detector is essential to optimize the DAMPE measurements, including charge, energy, and direction measurements for individual events. The trigger and data acquisition (DAQ) system are also tuned according to the on-orbit situation to optimize the detection efficiency. The purpose of this work is to document the on-orbit calibration procedures used by DAMPE and present the calibration results. An overview of the calibration is presented in Section 2. The calibration details on trigger, PSD, STK, BGO, and NUD are then described in Sections 3-7, respectively. The determination of the perimeter of the South Atlantic Anomaly (SAA) is introduced in Section 8. The measurement of the live time is described in Section 9. The detector alignments are given in Section 10. We summarize this work in section 11.

\section{Overview of the DAMPE calibration}

DAMPE is operating in a 500 km solar-synchronous orbit with an inclination of $\sim97^\circ$ and a period of about 95 minutes. During its on-orbit operation, calibration data are acquired in two different modes, namely, the ``dedicated-mode" for electronics linearity and pedestal calibration, and the ``continuous-mode" for science data-taking. The details of the DAQ of these two modes will be introduced in the next section. The DAQ system runs in the continuous-mode for most of the time. It switches to the dedicated-mode for 40 second pedestal calibration twice per orbit, at the latitude of about $+20$ deg in the terrestrial coordinate. The calibration data are also acquired in the dedicated-mode for electronics linearity measurements. The data acquired in the continuous-mode for nominal science data-taking are used to calibrate and monitor the detector's performance. In the latitude region between $-20$ degree and $+20$ degree, DAMPE enables the Minimum Ionization Particle (MIP) trigger logic \cite{Chang2017} {since the geomagnetic cut-off of cosmic rays is about 10 GV, for which the ionization energy loss rate, as described by the Bethe-Bloch formula, reaches the minimum.  
}

Table~\ref{tab-List} summarizes the DAMPE calibration and configuration settings. Here the configurations have been fixed before the launch, while the calibration settings can be modified during the operation when necessary.
Several kinds of calibration need to be done throughout the lifetime of the DAMPE mission, including 
the ``zero" data compression thresholds, the trigger thresholds, the calculation of detector gains, the MIP response and light asymmetry, the evaluation of the boundary of the South Atlantic Anomaly (SAA), the measurement of the live time, and the detector alignments.

Before the launch, intensive calibration had been carried out to understand and optimize the performance of DAMPE.
Dedicated experiments were carried out to study the temperature dependence of the pedestal, the gain and other electronics parameters. After the launch, all these kinds of calibration have been repeated using cosmic rays. Comparison with the pre-launch results has been made to verify the on-orbit performance of the detector.

\begin{table}[!htb]
\begin{small}
\begin{center}
\caption{List of DAMPE calibrations (calib) and configurations (config). Detail descriptions are given in corresponding sections.}
\label{tab-List}
\begin{tabular}{lllllll}
\hline
 Category & Parameter                              & Type (calib/config) & DAQ Freq.           & DAQ Mode            &Section & Comment                   \\
\hline
 PSD & pedestal                                        & calib                      & twice per orbit       & dedicated               & IV (A)   & at lat. of +20 deg.       \\
 PSD & zero-suppression threshold          & config                    & twice per orbit       & dedicated               &             & at lat. of +20 deg.       \\
 PSD & electronics linearity                       & calib                      & once a month        & dedicated               &             & not in SAA                  \\
 PSD & ratio of Dy5, Dy8                           & calib                      & continuous            & continuous             & IV (B)   &                                    \\
 PSD & MIP response and energy scales  & calib                      & continuous            & continuous             & IV (C)   &                                    \\
 PSD & light attenuation                             & calib                      & continuous            & continuous             & IV (D)   &                                    \\
 PSD & intra detector alignment	                 & calib                      & continuous            & continuous             & X (A)   &                                    \\
 STK & pedestal                                         & calib $\&$ config   & twice per orbit       & dedicated               & V (A)    & at lat. of +20 deg.       \\
 STK & threshold for cluster finding           & config                    & once a day            & dedicated	       & V (A)    & raw data mode           \\
 STK & noisy channels                               & config                    & continuous            & continuous             & V (A)    &                                    \\
 STK & electronics linearity                        & calib                      & once per month     & dedicated               & V (B)    & not in SAA \\
 STK & MIP response and energy scales  & calib                       & continuous             & continuous            & V (B)     & \\
 STK & intra detector alignment	                 & calib                      & continuous             & continuous             & X (B)    & \\
 BGO & pedestal                                        & calib                      & twice per orbit         & dedicated              & VI (A)   &at lat. of +20 deg. \\
 BGO & MIP response and energy scales  & calib                       & twice per orbit        & continuous            & VI (B)   & MIP trigger pattern enabled  \\
          &                                                       &                                &                               &                              &              & in lat. $(-20, +20)$ deg. region \\
 BGO & zero-suppression threshold          & config                     & twice per orbit        & dedicated              &             & at lat. of +20 deg.\\
 BGO & electronics linearity                       & calib                       & once a month         & dedicated              &             & not in SAA \\
 BGO & ratio of Dy2, Dy5, Dy8                   & calib                      & continuous              & continuous            & VI (C)   & \\ 	
 BGO & light attenuation                             & calib                       & continuous             & continuous            & VI (D)   &\\ 	
 BGO & trigger threshold                            & calib                       & continuous             & continuous            & VI (E)    & \\ 	
 NUD & integ. gate open to trigger delay   & config                      & continuous             & continuous           & VII         &\\ 	
 NUD & integ. gate close to trigger delay   & config                     & continuous             & continuous            &              & \\
 Others & SAA polygon                              & calib                        & continuous             & continuous            & VIII        & \\
 Others & DAMPE Live Time                      & calib                        & continuous             & continuous            & IX        & \\
 Others & DAMPE boresight                      & calib                        & continuous	          & continuous           & X (C)     & \\ 	
\hline
\end{tabular}
\end{center}
\end{small}
\end{table}

\section{Trigger and DAQ}

The trigger and DAQ system 
controls the data flow from the detector to the flash mass memory \cite{Chang2017}. The DAMPE DAQ system packages the zero-suppressed science data in the front-end electronics. In different acquisition modes, DAMPE uses different trigger logic patterns generated by the level one hardware trigger logic (L0).


Hit signals from the BGO calorimeter are used to generate the trigger \cite{Chang2017}.
Different hit thresholds are set for different readout channels in the BGO, resulting in different energy threshold for the trigger decision.
In the continuous-mode, the trigger logic is configured to acquire the science data for the MIP calibration and the detection of high energy electrons and gamma-rays. In this mode, the dead-time is fixed to be about 3 ms. When receiving the trigger signal, the front-end electronics of the detector digitizes the analog signals
and package the science data with zero-suppression. {Pedestal and electronics linearity calibrations are carried out in the dedicated-mode, in which a periodic trigger signal is used and the trigger rate is reconfigured.} For these calibrations, the trigger is configured to produce a specified number of events at a particular rate. For the pedestal calibration, the DAQ collects raw data of the detector's pedestal and transfers them to the ground. The DAMPE data center analyzes the pedestals on the ground and decides whether it is necessary to update the configuration of the zero-suppression thresholds. For the electronics linearity calibration, the front-end electronics boards will receive the trigger signal and then inject a known, programmable amount of charge to the pre-amplifiers which are also connected to the detector. The procedure is repeated with different charge injection to cover the whole dynamic range.



\section{PSD calibration}
\label{sec:psdcali}

As the top-most sub-detector of DAMPE, the PSD is designed to measure the charge of cosmic-ray nuclei with atomic number $Z$ from 1 to 26 and to reject the charged-particle background for gamma-ray detection. The PSD has an orthogonal double-layer configuration, with 41 detector modules in each layer. Each module is composed of a plastic scintillator bar and two photomultiplier tubes (PMTs) mounted on each end of the bar. The size of the PSD bar is 884 mm $\times$ 28 mm $\times$ 10 mm (884 mm $\times$ 25 mm $\times$ 10 mm for the bars at edges). To cover a wide dynamic range of energy measurement, signals from dynode 5 (Dy5) and dynode 8 (Dy8) of the PMTs are extracted. To minimize the influence of the temperature variation on the PSD performance, an active temperature control strategy is adopted \cite{YuYH2017}. After more than one year's operation in space, it is found that the temperature variation of the PSD is less than 1$^\circ$C. More detailed information about the PSD design can be found in Refs.~\cite{YuYH2017,Zhouyong2016range,Zhouyong2016pmt}.

The PSD on-orbit calibration includes: (A) the pedestal calibration, (B) the PMT dynode calibration, (C) the MIP response calibration, (D) the energy reconstruction, and (E) the light attenuation calibration.

\subsection{Pedestal calibration}
\begin{figure}
 \begin{center}
    \vspace{-2mm}
    \centering
    \includegraphics[scale=0.63]{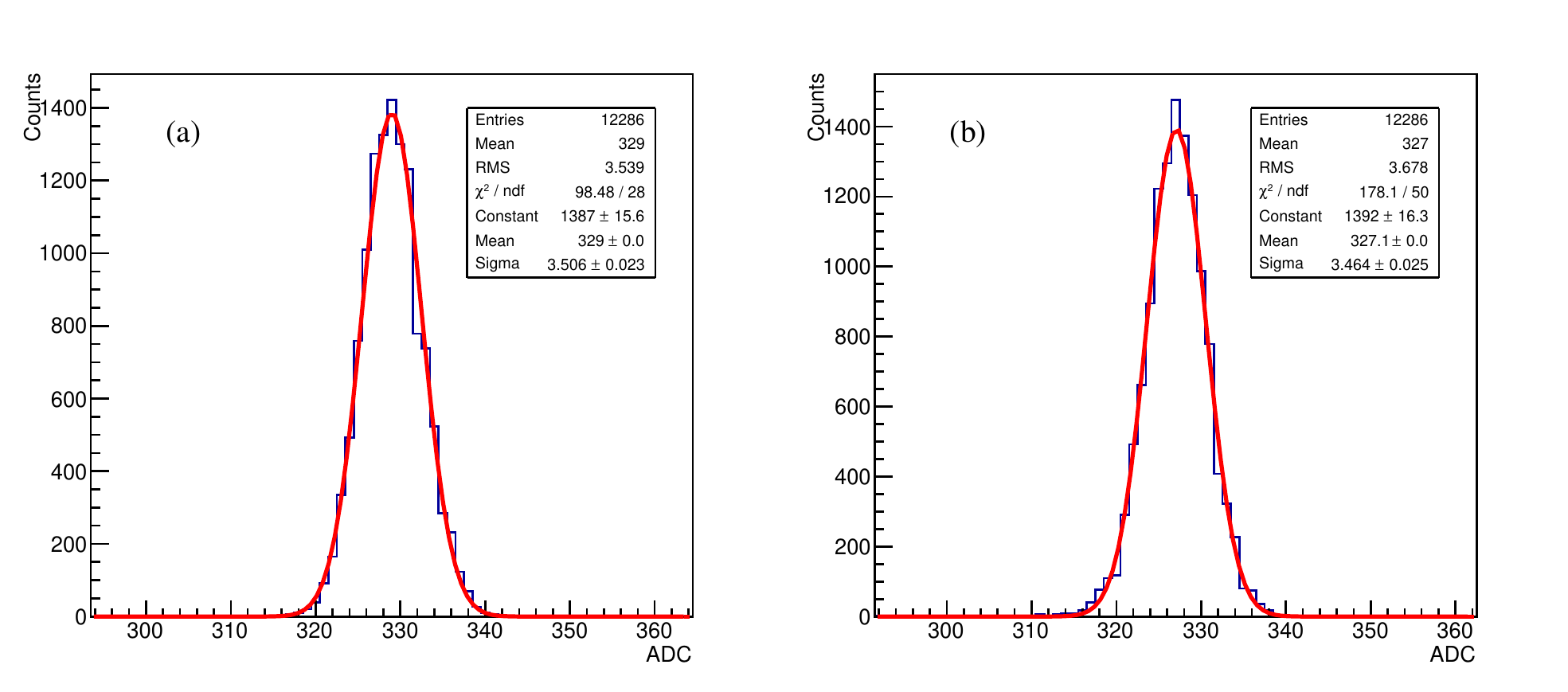}
    \caption{Typical PSD pedestal distributions of Dy5 (a) and Dy8 (b) of a PMT, together with Gaussian fittings.}
     \label{fig:ped}
 \end{center}
\end{figure}

Pedestals are offset voltages present at the Analog-to-Digital Converter (ADC) inputs in the readouts. The pedestals of the PSD are measured twice per orbit using a 100 Hz periodic trigger, which provides about 6000 random events per orbit. Fig.~\ref{fig:ped} shows the typical pedestal distributions of Dy5 (left panel) and Dy8 (right panel) from one PMT. Solid lines show the fitted Gaussian functions. The fitted variances ($\sigma$) of the pedestal distributions are about 3.5 ADC. The pedestals are updated daily during the data processing.

\begin{figure}
 \begin{center}
    \vspace{-2mm}
    \centering
   \includegraphics[scale=0.6]{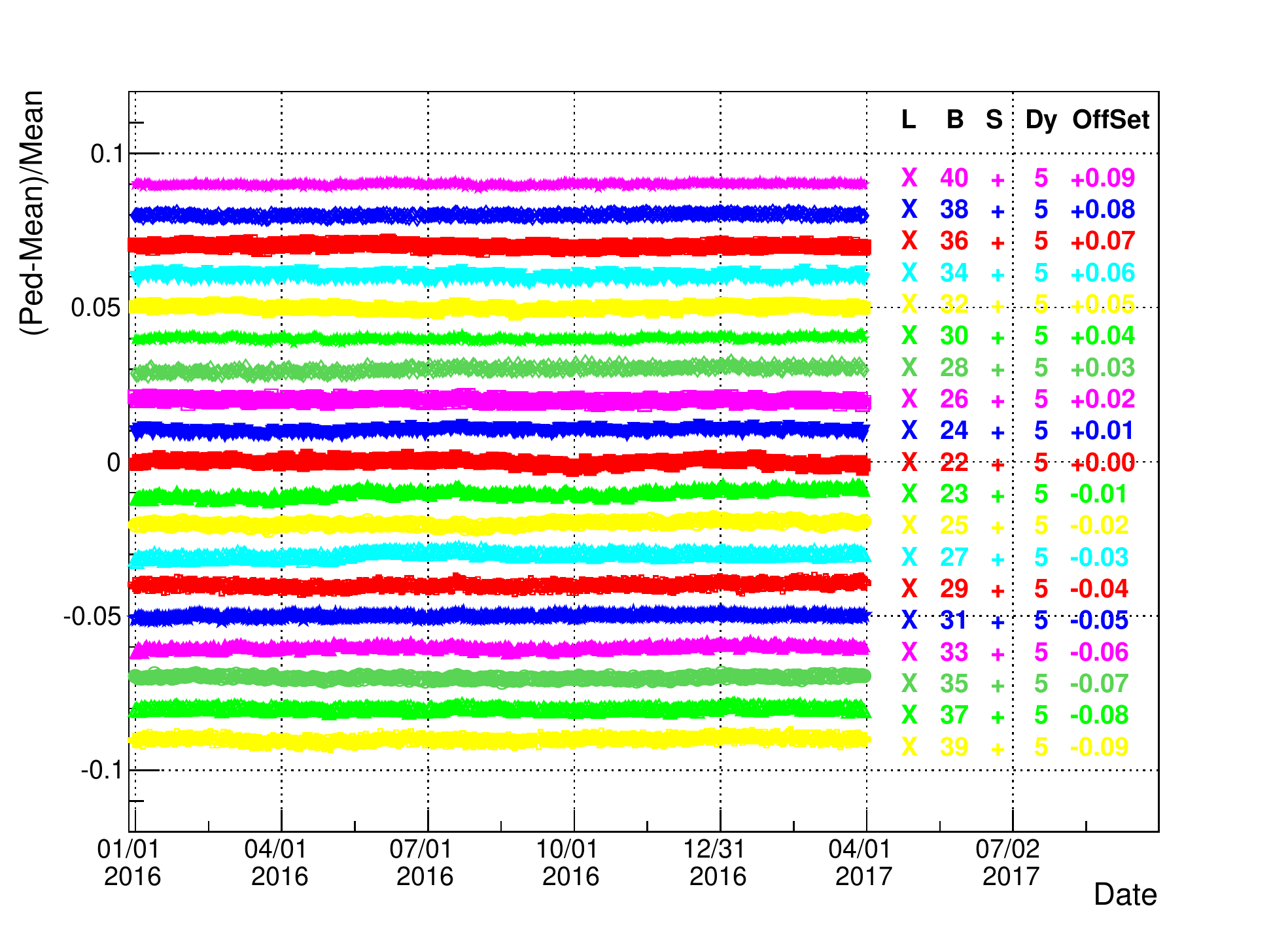}
    \caption{Variations of PSD pedestals in 15 months. The corresponding layer (L), bar (B), side (S), and dynode number are listed on the right side. Offsets are adopted to each curve for clarity of demonstration.}
     \label{fig:pedDate}
 \end{center}
\end{figure}

In order to characterize the stability of the pedestal, we define a pedestal variation quality as $(Ped - Mean)/Mean$, where $Ped$ is the one-day average value of the pedestal, and $Mean$ is the long time average. In Fig.~\ref{fig:pedDate}, the pedestal stabilities of the channels associated with Dy5 of the positive side PMTs 22-40 of the top layer are presented \cite{Chang2017}. We can see that the pedestals of the PSD are very stable and the overall variation is less than 0.1\% during the first year's operation.

\subsection{PMT dynode calibration}

\begin{figure}
\begin{center}
    \centering
\includegraphics[width=0.8\textwidth]{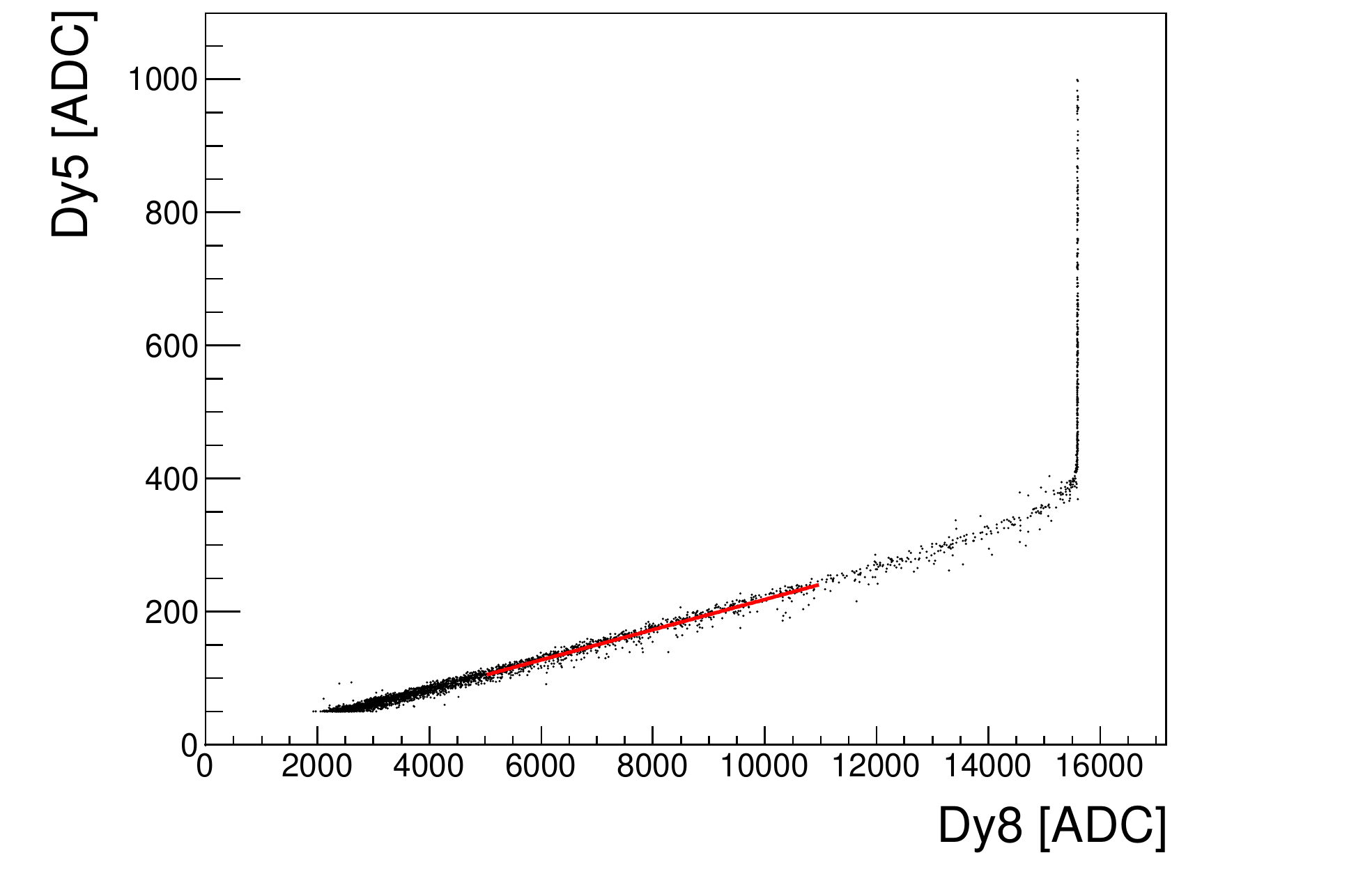}
    \caption{ The correlation between Dy5 (vertical axis) and Dy8 (horizontal axis) for a certain PMT. The red line shows the linear fitting.}
    \label{fig:PsdDy58}
\end{center}
\end{figure}

After subtracting the pedestals, ADCs of Dy5 and Dy8 of a PMT show a linear correlation up to 14000 ADC of Dy8 { (see FIG. \ref{fig:PsdDy58}).%
We use a linear function to fit such a correlation. The points which are 5$\sigma$ away from the correlation band are excluded in the fit.
The correlation parameters, the slope $K$ and the intersection $C$, are extracted for each PMT. The final ADC of each PMT is calculated by}
\begin{equation}
    ADC=
    \begin{cases}
        {\text{Dy8}} &\mbox{(Dy8 $\leq11000$)}\\
        K{\times}\text{Dy5}+C &\mbox{(Dy8 $>11000)$}
    \end{cases}.
    \label{eq:dy58}
\end{equation}

\subsection{MIP response and energy reconstruction}

\begin{figure}
\begin{center}
    \centering
    \includegraphics[scale=0.6]{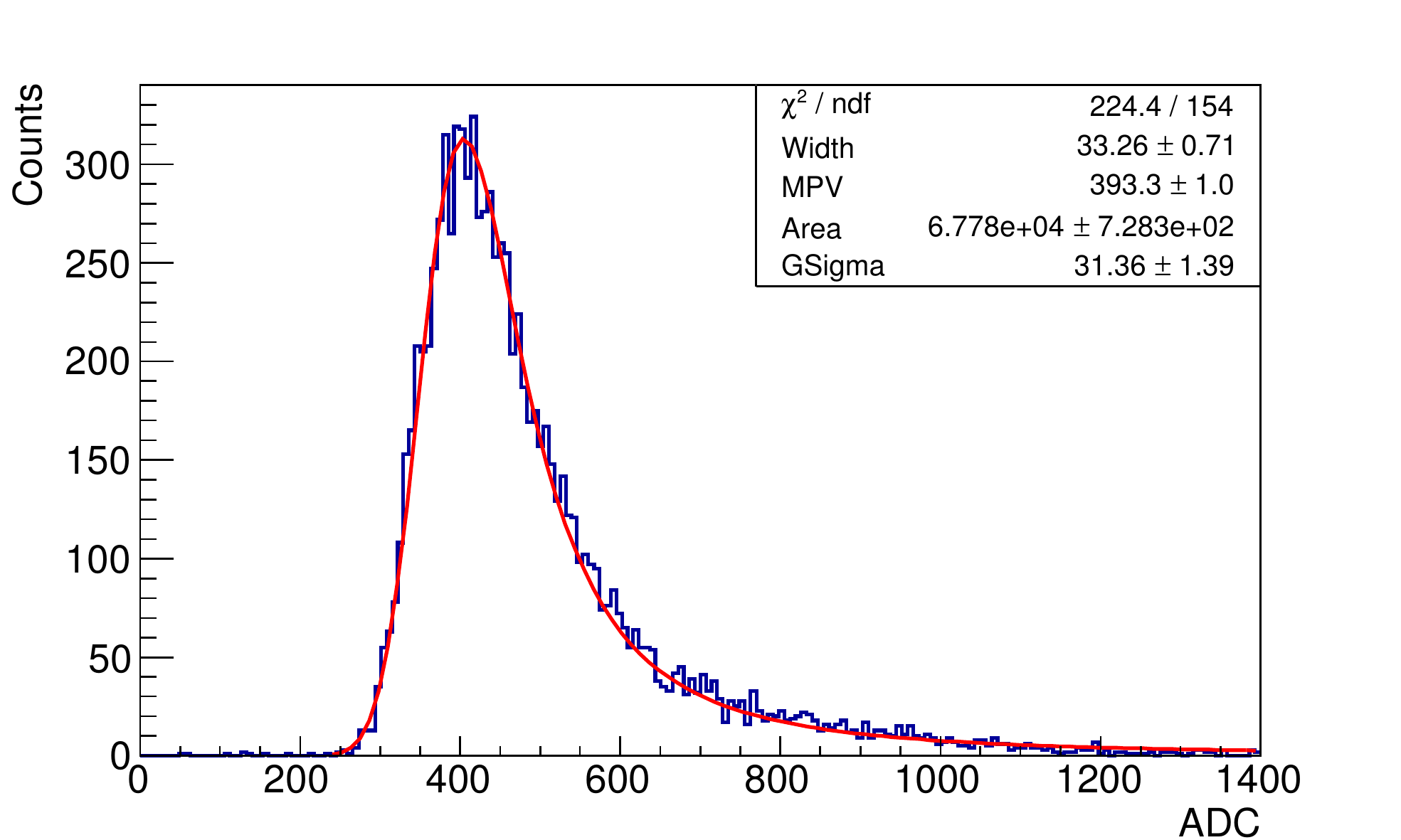}
    \caption{Typical ADC distribution of MIPs (histogram), together with the fitting curve with a Gaussian convolving Landau function (solid curve).}
    \label{fig:mipsadc}
\end{center}
\end{figure}

\begin{figure}
\begin{center}
    \centering
    \includegraphics[scale=0.6]{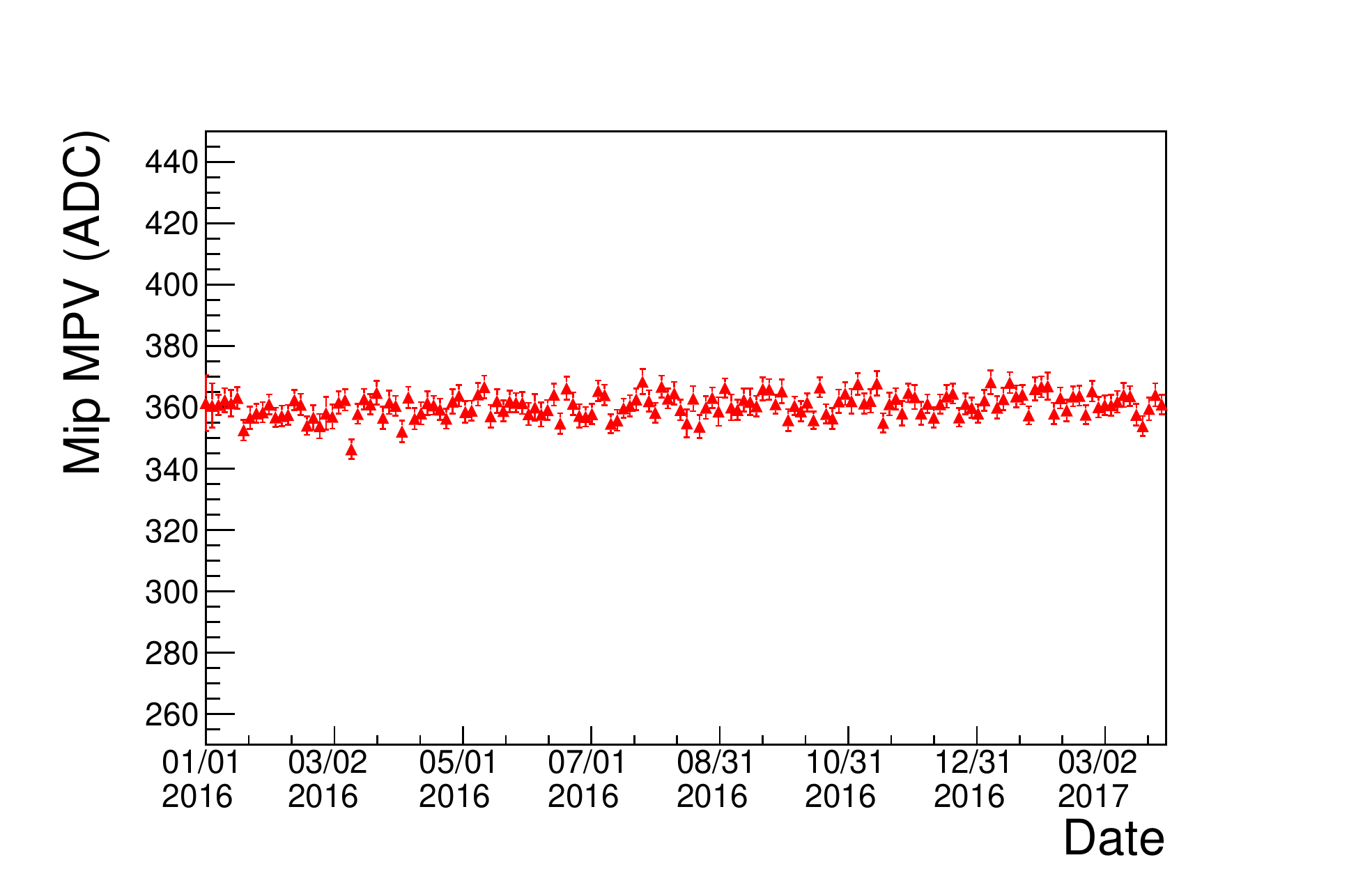}
    \caption{MIP response as a function of time.}
    \label{fig:MipvsTime}
\end{center}
\end{figure}

Proton MIP events are used to calibrate the response of the energy deposition in the PSD. The proton MIP sample is selected using the BGO information. We refer the readers to Ref. \cite{WangYP2017} for the selection algorithm. Fig.~\ref{fig:mipsadc} shows the typical ADC distribution of a PMT for MIP events. The distribution can be fitted by a Gaussian convolving Landau distribution, as shown by the red solid curve in the figure. The most probable value (MPV) and the width of the Landau function are obtained. A combined MIP response of a PSD bar ($MPV_C$) is calculated as
\begin{equation}\label{eq:mipadc}
  MPV_{C}=\sqrt{MPV_L{\times}MPV_R},
\end{equation}
using the MPV values obtained for the left-side ($MPV_L$) and right-side ($MPV_R$) readouts of such a bar. The MIP response of the PSD is very stable, as illustrated in Fig.~\ref{fig:MipvsTime}.

The obtained MPV values (i.e., $MPV_L$, $MPV_R$ and $MPV_C$) are then used to reconstruct the energy deposition of an incident charged particle.
According to the Bethe formula~\cite{PDG2016}, the energy deposition of proton MIP events in the PSD can be used as a unit to measure the energy deposition of other cosmic ray particles. Considering that the typical energy loss of proton MIPs in the PSD is about 2 MeV/cm and that the thickness of a PSD bar is 1 cm, the energy deposition ($E_{L/R/C}$) measured by the left-side/right-side/combined readouts of a PSD bar can be given as
\begin{equation}\label{Energy_rec}
E_{L/R/C} =  \frac{ADC_{L/R/C}}{MPV_{L/R/C}}\times2~\text{MeV}.
\end{equation}

\subsection{Light attenuation calibration}
\begin{figure}
\begin{center}
    \centering
    \includegraphics[scale=0.6]{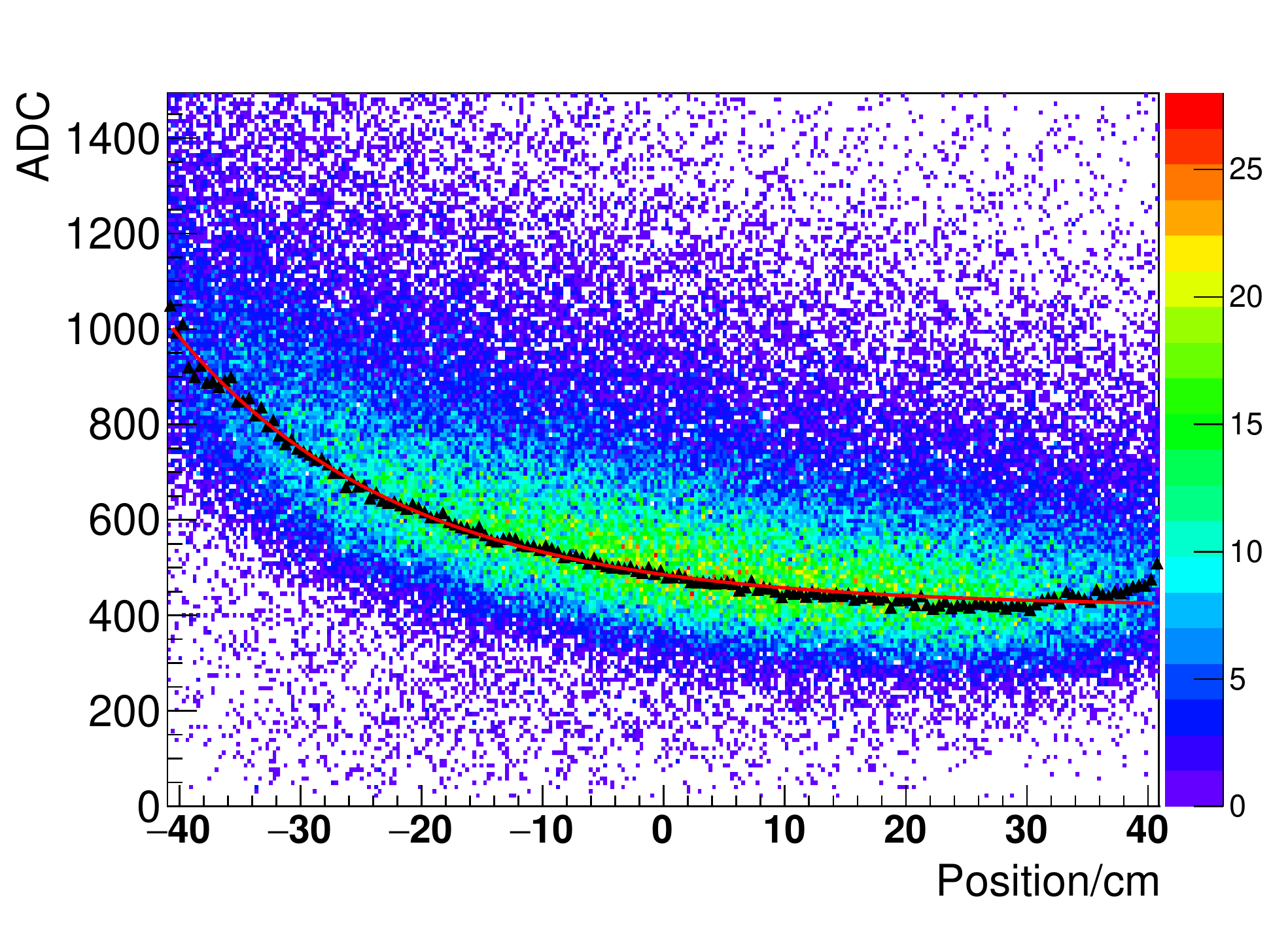}\\
    \caption{ADC values (vertical axis) of MIPs versus the hit positions (horizontal axis). The triangles represent the MPV values, and the red solid line is the best fit function of the MPV-position relation.}
    \label{fig:attcomp}
\end{center}
\end{figure}

The scintillation light will get attenuated during the propagation in the PSD bar. Therefore, the measured energy needs to be properly corrected.
The attenuation correction depends on the hit position and hence the track of the incident particle. To ensure that a right track is selected, the MIP events with only one STK track are used. The STK track is back-traced to determine the hit position in the PSD. The light attenuation behavior for each side of the PSD bar is obtained via the obtained hit position along the strip and its measured energy. Fig.~\ref{fig:attcomp} shows a typical scattering plot of hit positions (horizontal axis) and path-length corrected energies (vertical axis) for one side of a PSD bar. 
The MPV value of each position bin (marked as triangle in Fig.~\ref{fig:attcomp}) is obtained by fitting the corresponding energy distribution with a Landau function.
The attenuation function for each side of every PSD bar is then obtained by fitting the MPV values versus the hit positions with the following empirical function
\begin{equation}\label{Eq:PsdAtt}
    A(x) =  C_0e^{-x/\lambda}+C_1x+C_2x^2+C_3x^3,
\end{equation}
where $\lambda,~C_0,~C_1,~C_2,~C_3$ are free parameters. The combined attenuation function is defined as $A_C(x)=\sqrt{A_L(x){\times}A_R(x)}$, with subscriptions ``L'' and ``R'' being the left and right side.

\subsection{Charge reconstruction}

The charge of an incident particle can be derived from the reconstructed energy deposition and the attenuation function, as
\begin{equation}\label{Eq:Qrec}
Q_{L/R/C}^{\rm rec}=\sqrt{\frac{E_{L/R/C}}{A_{L/R/C}(x)}{\times}\frac{\text{D}}{s}},
\end{equation}
where $Q_{L/R/C}^{\rm rec}$ is the reconstructed charge by the left/right/combined readout, $E_{L/R/C}$ is the reconstructed energy, $A_{L/R/C}(x)$ is the attenuation as a function of hit position $x$, $s$ is the path length of the particle in one PSD bar given by the selected track, and $\text{D}=10$ mm is the thickness of the PSD bar.

\begin{figure}
    \centering
    \includegraphics[width=0.8\textwidth]{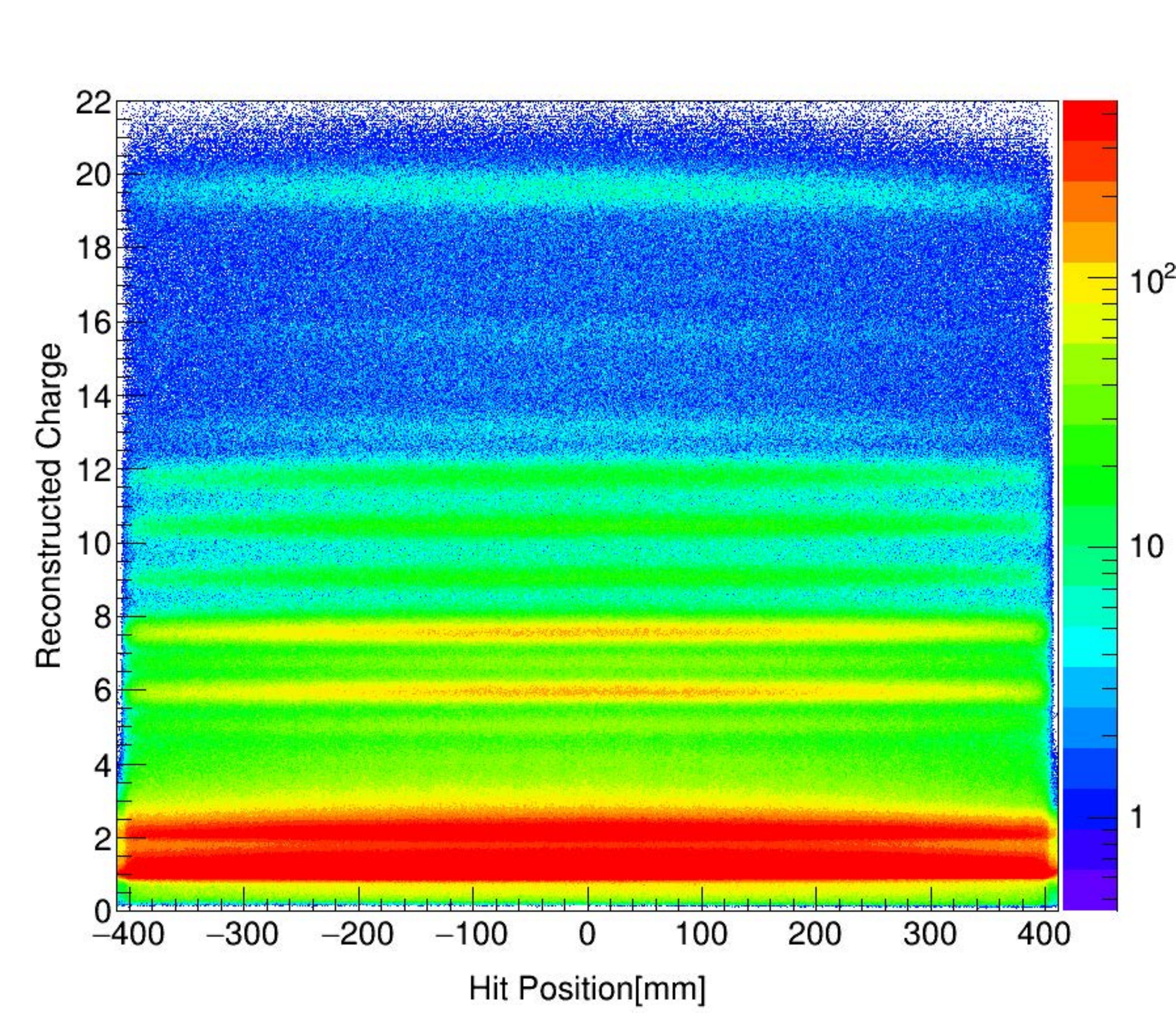}
    \caption{ Reconstructed charge (vertical axis) versus the hit position (horizontal axis).}
    \label{fig:qpos}
\end{figure}

\begin{figure}
    \centering
    \includegraphics[width=0.8\textwidth]{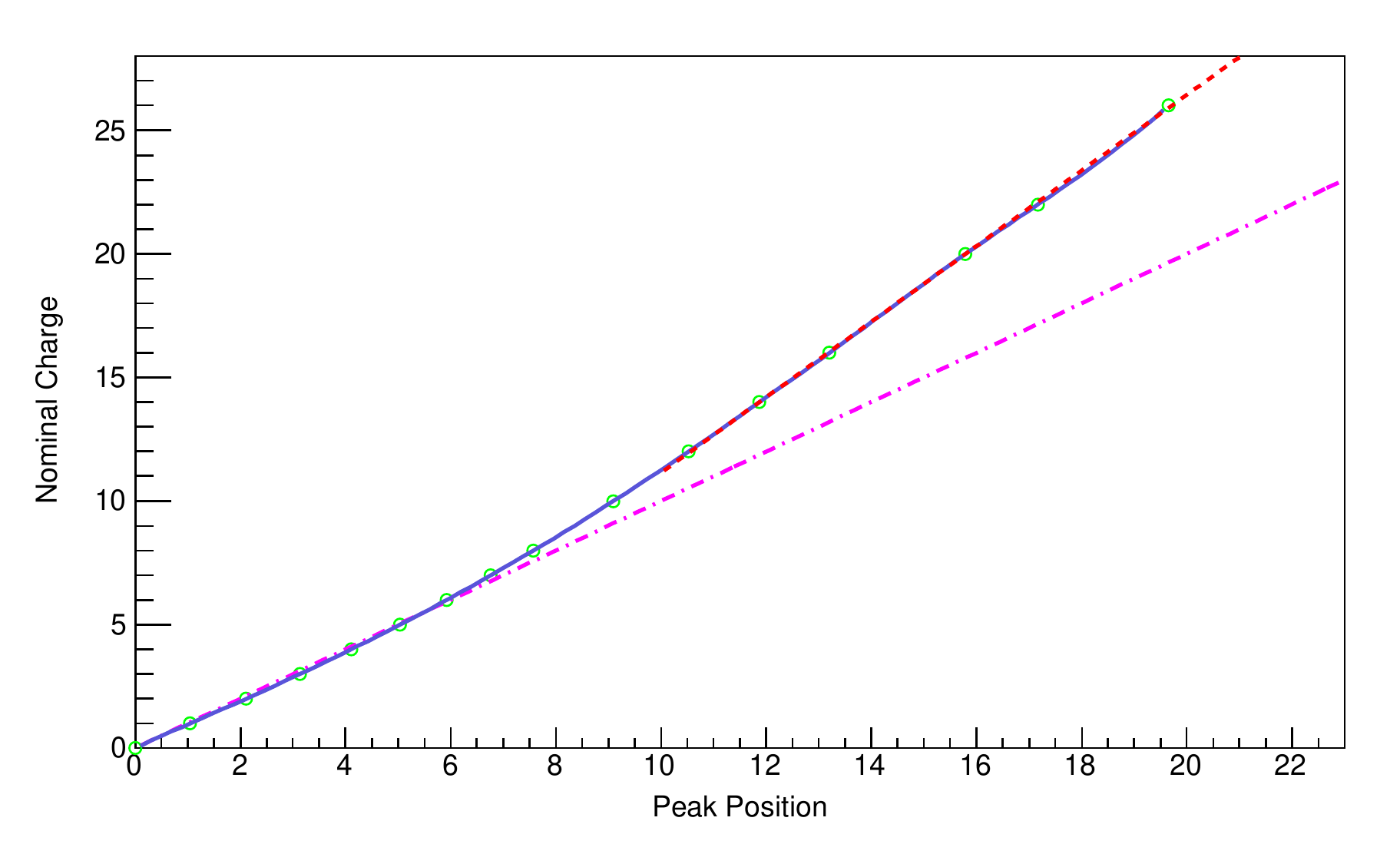}
    \caption{ Peak positions in a reconstructed charge spectrum versus the nominal charge numbers (open circles). The solid line represents the third-order spline function fitting, the dashed line is the fitted linear function, and the dash-dotted line represents $y=x$.}
    \label{fig:Que}
\end{figure}

{ Fig.~\ref{fig:qpos} shows the reconstructed charge of particles for different hit positions. We can find major classes of cosmic ray nuclei, such as protons, Helium, Carbon (with $Q^{\rm rec}\approx6$), Oxygen (with $Q^{\rm rec}\approx7.5$), and Iron (with $Q^{\rm rec}\approx19$). The reconstructed charge for high-$Z$ particles deviates from their actual charge due to the quenching effect.
In order to correct such an effect, we fit the peaks of reconstructed charge distributions (left/right/combined sides) with a series of Gaussian functions. Through comparing the best-fitting values of each peak with the actual charge of corresponding nuclei groups, a renormalization factor of the reconstructed charge can be obtained. Fig.~\ref{fig:Que} shows the obtained peak positions by the reconstructed charge versus the nominal charge for one PSD bar.
We use a third-order spline function to fit the relation of reconstructed peak values and the corresponding nominal values, for each kind of charge measurement (left/right/combined side) of each PSD bar.
For cosmic nuclei beyond Iron nuclei (with $Q^{\rm rec}\sim19$), a linear extrapolation is used based on the linear function fitted with data with $Q^{\rm rec}>10$.
For each reconstructed charge, a quenching-effect corrected charge $Q_{L/R/C}$ can be obtained via}
\begin{equation}\label{Eq:Qcor}
    Q_{L/R/C} = f_{L/R/C}(Q_{L/R/C}^{\rm rec}),
\end{equation}
where $f_{L/R/C}(Q^{\rm rec}_{L/R/C})$ is the best-fitting third-order spline function for $Q^{\rm rec}<19$ and the linear function for $Q^{\rm rec}\geq 19$. After this correction, the charge spectra measured by different PSD bars can be directly combined.

\begin{figure}
\begin{center}
    \centering
    \includegraphics[width=0.8\textwidth]{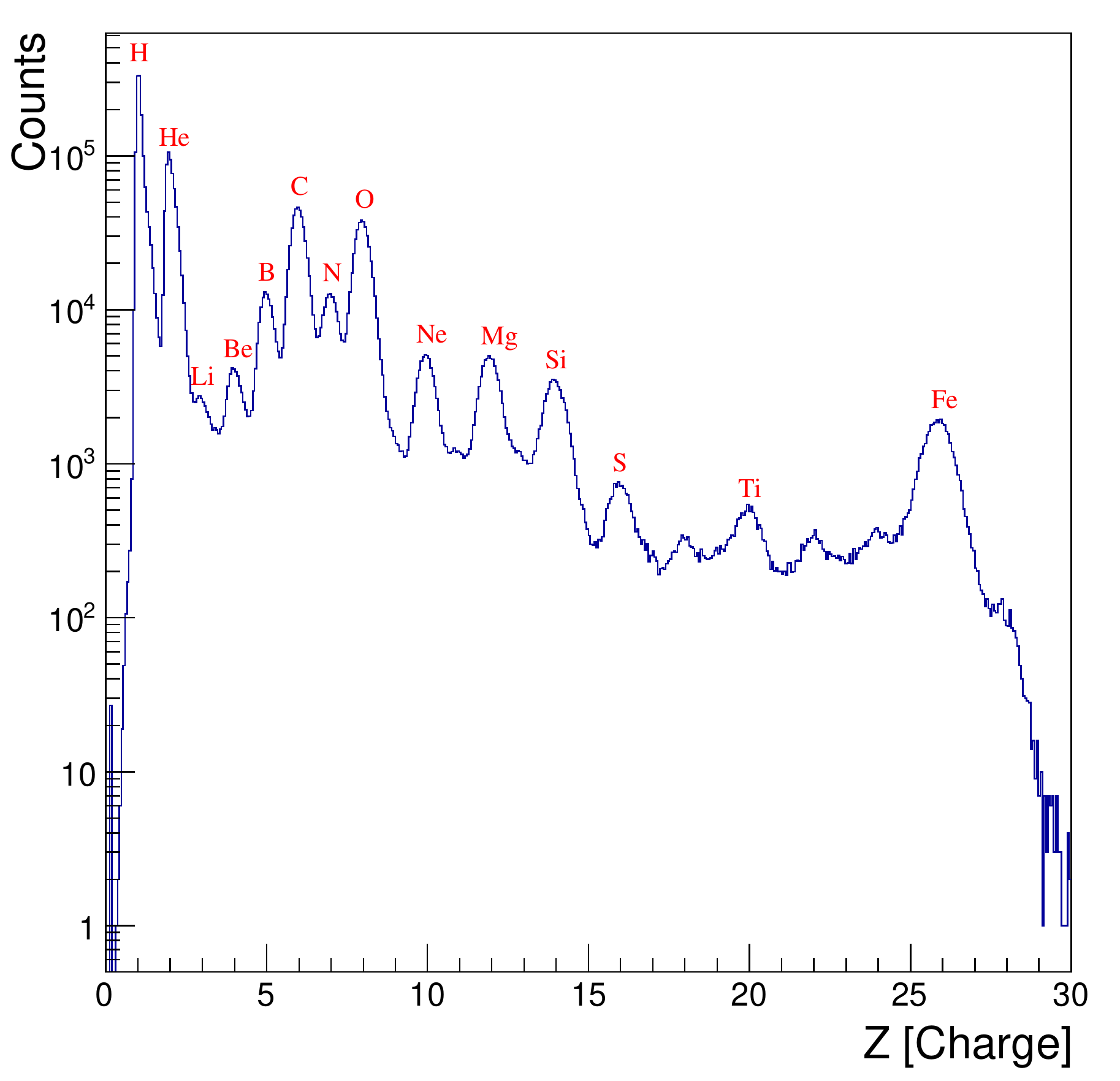}\\
    \caption{Charge spectrum with 15 days of the flight data.}
    \label{fig:chrg}
\end{center}
\end{figure}

Fig.\ref{fig:chrg} shows the final charge spectrum after the quenching correction from the top layer of the PSD with 15 days of the flight data. The charge peaks of the major cosmic ray nuclei are clearly visible. A more detailed charge reconstruction can be found in Ref.~\cite{DongTK2018}.

\section{STK calibration}

The STK is the second from top sub-detector of DAMPE,  designed to reconstruct the trajectories of charged particles, measure the absolute ion charge (Z) of cosmic-rays and pinpoint the incoming direction of detected photons. The STK consists of six tracking double layers providing  six independent $x$/$y$ particle coordinate measurements. The layers are physically mounted onto 7 support trays. To promote the photon conversions into $e^\pm$-pairs, three tungsten layers are inserted after the first, second and third layers respectively. Each tracking layer is composed of 16 modules, called ladders, read in groups of 24 by 8 data acquisition boards~\cite{Zhang:2016hth}. The ladders are formed by 4 single-sided AC-coupled silicon micro-strip sensors, daisy-chained via micro-wire bonds.  There are  384 readout channels per ladder, read by 6 VA ASIC chips. Further details about the STK design and construction can be found in~\cite{Azzarello:2016trx}. In this section, we describe the channel pedestal and noise calibration procedure and the VA calibration.  The STK silicon sensor alignment procedure is described in Section~\ref{sec:stk_alignment}.

\newcommand{\stkjanuaryfirst}{January 1 2016}
\newcommand{\stkapriltwentysix}{April 26 2016}
\newcommand{\stkaugustseventeen}{August 17 2016}

\subsection{Pedestal and noisy channels}
\begin{figure}[!htb]
	\centering
	\includegraphics[scale=0.5]{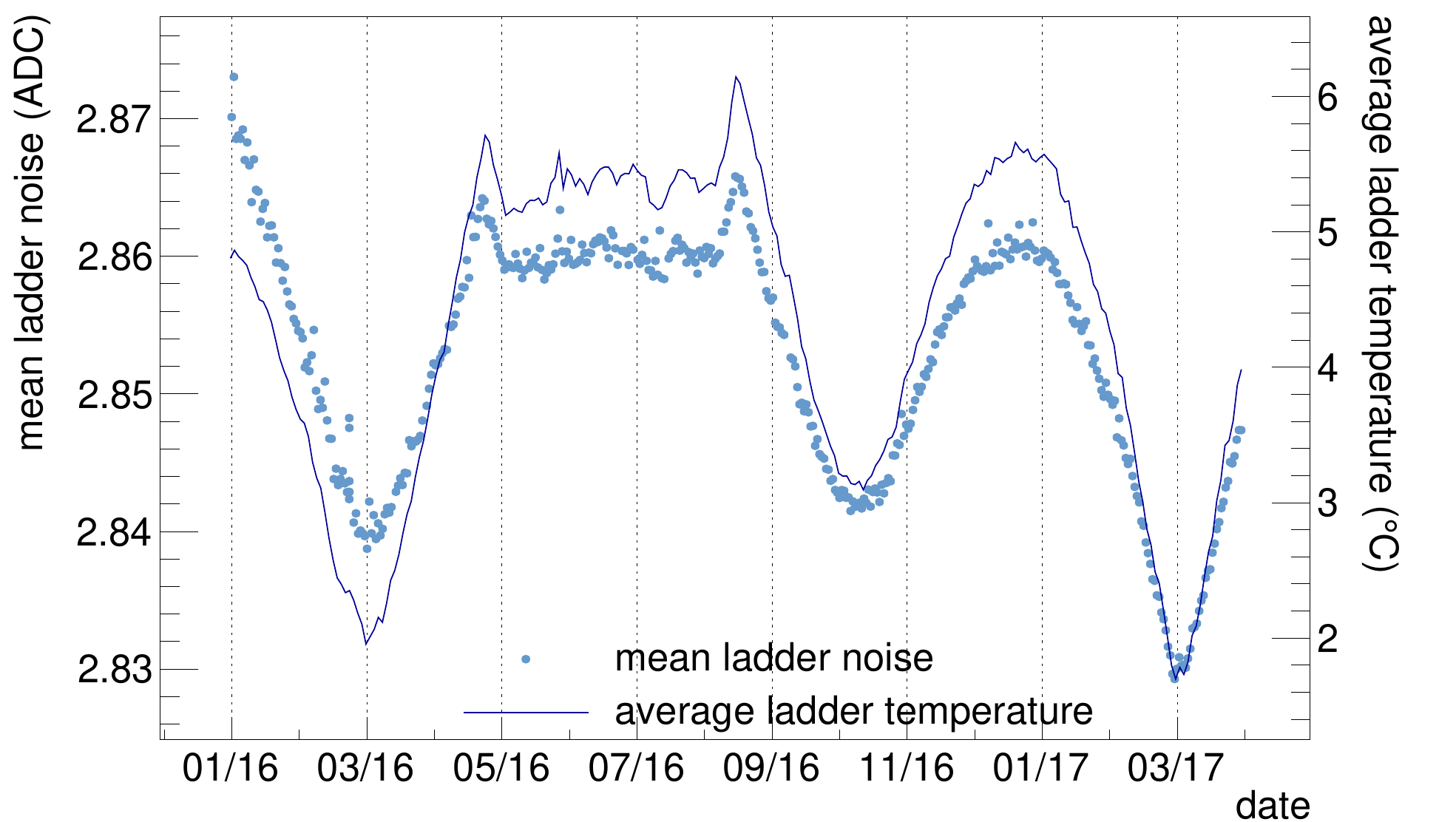}	
	\caption{All ladder strip noise evolution since \stkjanuaryfirst~ (light blue points).
	The noise variation over more than one year is less than 1.4\%. The temperature varies from 1.8 $^\circ$C and 6.2 $^\circ$C, the daily variation is less than 0.15 $^\circ$C.}
	\label{fig:agvnoisevstime.pdf}
\end{figure}

The STK has a total of 73728 channels that are constantly monitored and it is the DAMPE's sub detector with the highest number of channels. Calibration procedure is needed to compute pedestal and noise level of each single channel of the detector. In order to get the best data quality, calibration must be performed with a random trigger, i.e. non correlated with particles crossing the detector, and in regions with low particle flux. Therefore, the raw data for the offline pedestal calibration is acquired every day. Pedestals, noise before common noise subtraction ($\sigma_{raw}$) as well as noise after common noise subtraction ($\sigma$), for each channel of each ladder, are computed~\cite{gallo_icrc15}. Also, the amount of channels with $\sigma\leq5$, with $5<\sigma\leq10$ and with $\sigma>10$ is computed. Fig.~\ref{fig:agvnoisevstime.pdf} shows the average ladder noise and ladder temperature evolution since \stkjanuaryfirst. The most important noise source is the strip leakage current, which is dependent on the silicon temperature. Every front-end electronics board is equipped with two temperature sensors. Computing the average temperature of all the ladders of the STK (considering all the 384 temperature sensors), it is possible to compare the noise evolution of the STK with the average ladder temperature, as shown in Fig.~\ref{fig:agvnoisevstime.pdf}. The average temperature, since \stkjanuaryfirst~ fluctuates between 1.8 $^\circ$C and 6.2 $^\circ$C, and the daily variation is less than 0.15$^\circ$C per day. This temperature variation corresponds to a noise variation over more than one year of about 0.04 ADC counts (less than 1.4\% of the average noise value) which does not have an impact on the computation of the MIPs signal which is on average 50 ADC counts. 

\begin{figure}[!htb]
	\centering
	\includegraphics[scale=0.5]{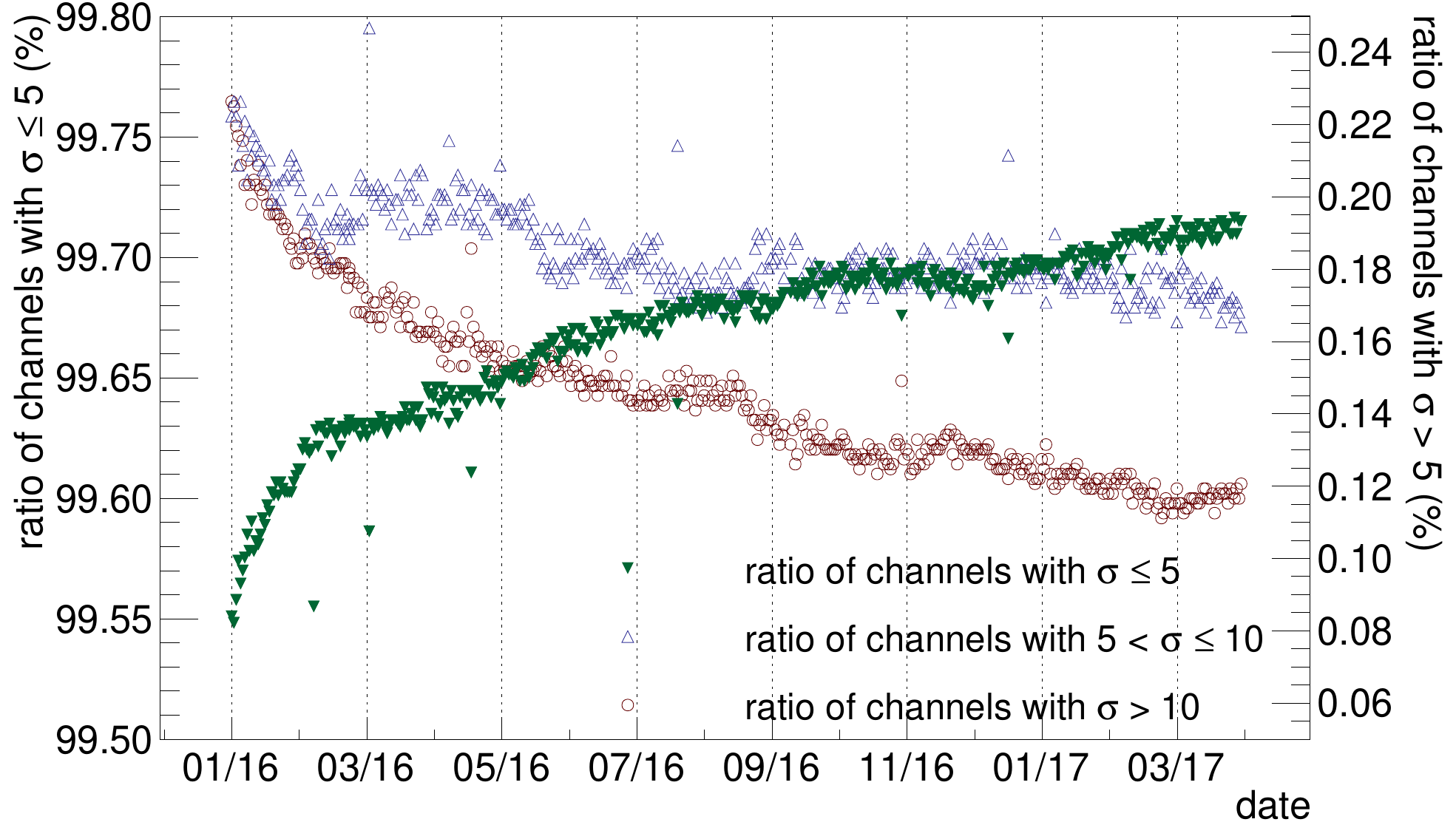}	
	\caption{Ratio of STK channels  with $\sigma\leq5.0$ (green), with $5.0<\sigma\leq10.0$ (blue) and with  $\sigma > 10.0$  (red) since the \stkjanuaryfirst.
	The percentage of channels with $\sigma<5.0$ increases with time, indicating that noisy channels progressively become good.}
	\label{fig:propnoisevstime.pdf}
\end{figure}

Here, a channel is considered as noisy if $\sigma>5.0$. Fig.~\ref{fig:propnoisevstime.pdf} shows the ratio of STK channels with $\sigma\leq5.0$,  with $5.0<\sigma\leq10.0$ and with $\sigma > 10.0$ since the \stkjanuaryfirst. The percentage of channels with $\sigma\leq5.0$ increases with time, indicating that noisy channels progressively become good again. At the same time, the percentage of channels with $5.0<\sigma\leq10.0$ is stable since July 2016, while the percentage of channels with  $\sigma\leq5.0$ continues to decrease.
This is due to the fact that very noisy channels (with $\sigma>10$) also improve thanks to the progressive stabilization of silicon sensors. Namely, some trapped impurities and dust leading to noisy channels get cleaned up with outgassing in space.

\begin{figure}[!htb]
	\centering
	\includegraphics[scale=0.5]{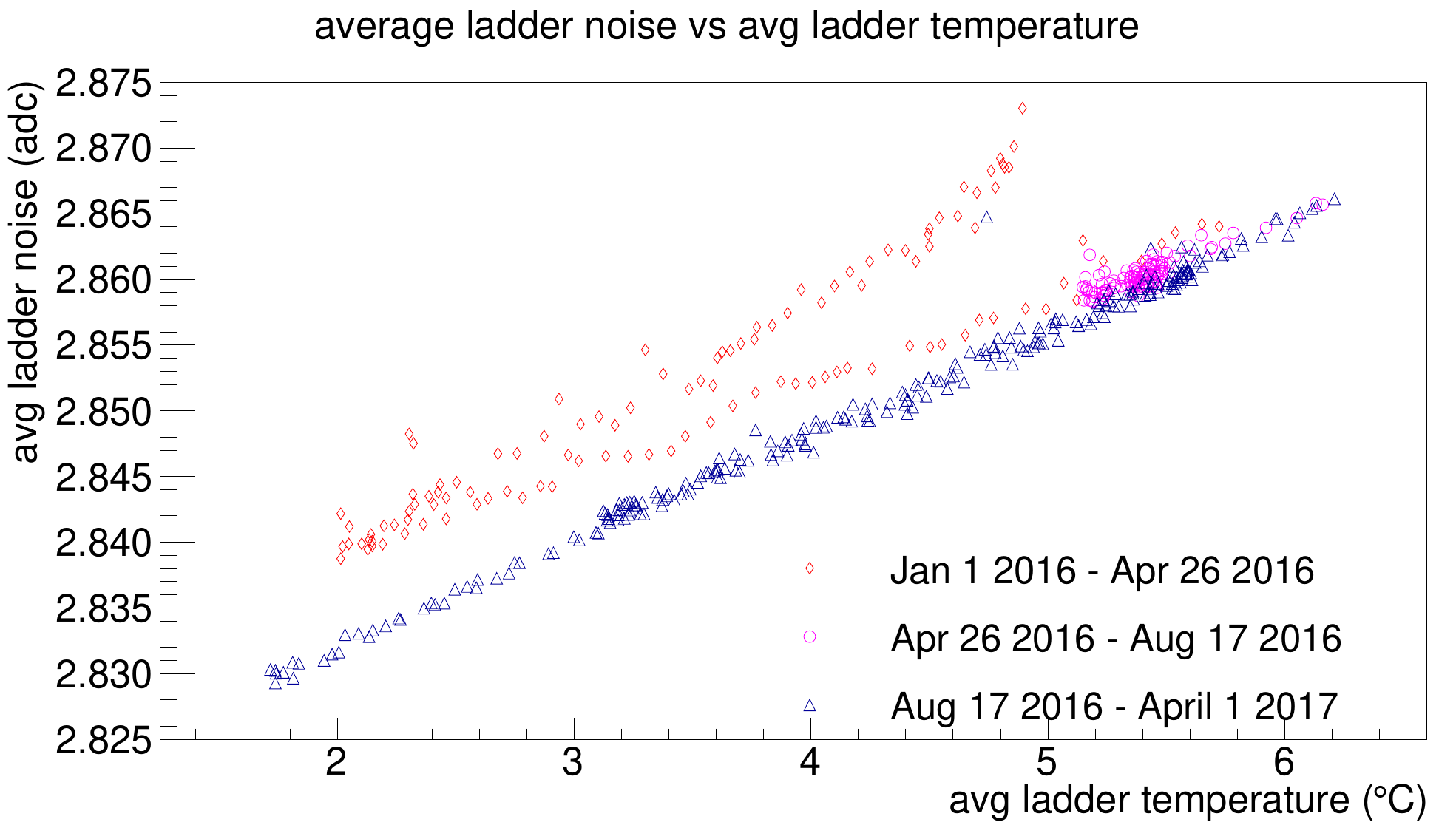}
	\caption{Average noise of the 192 ladders of the STK, with respect to the average ladder temperature. The red markers correspond to the \stkjanuaryfirst~--\stkapriltwentysix~ period. The magenta markers correspond to the \stkapriltwentysix~ -- \stkaugustseventeen ~ period. The blue markers correspond to the measurements since \stkaugustseventeen.}
	\label{fig:avgnoisevstemp.pdf}
\end{figure}
\begin{figure}[!htb]
	\centering
	\includegraphics[scale=0.5]{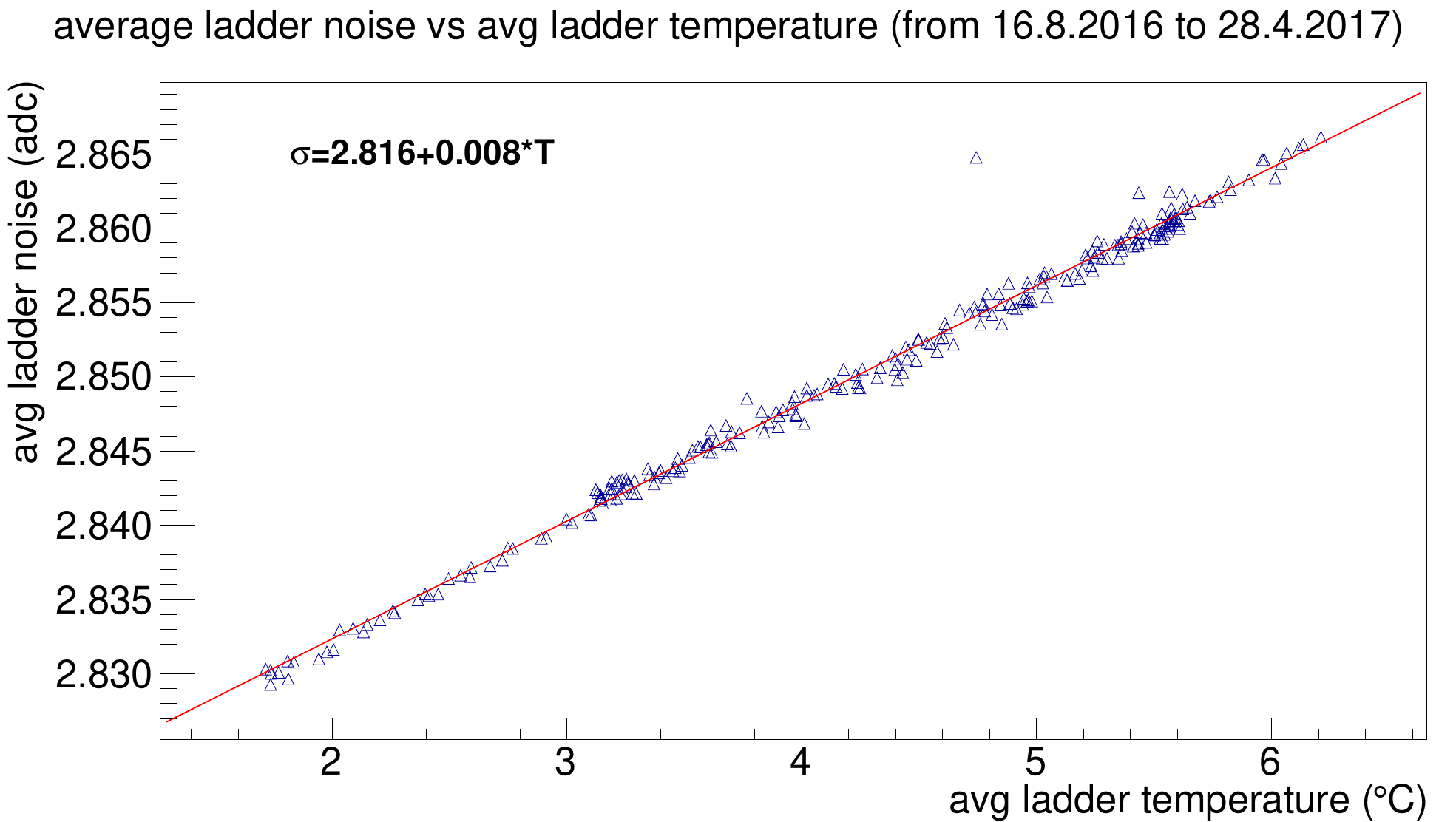}
	\caption{Ladder average noise vs ladder average temperature, since \stkaugustseventeen. The distribution is linear and the fit results shows that the noise varies of less 0.008 ADC per degree Celsius.}
	\label{fig:avgnoisevstempfit.pdf}
\end{figure}

The sensors measure the temperature of the front-end electronics, close to the ASICS which are the main heat source of the ladder. Thus the silicon temperature is lower of a few degrees than the measured temperature. Still, it is possible to examine the average ladder noise vs. the average ladder temperature, as shown in Fig.~\ref{fig:avgnoisevstemp.pdf}. There, the measurements are grouped into three time periods: \stkjanuaryfirst~-- \stkapriltwentysix, \stkapriltwentysix~--\stkaugustseventeen~and the period since~\stkaugustseventeen.
During the first period, noisiest channels improve rapidly with time, thus the reduction of the average noise, and the convergence towards the baseline (blue) distribution. During the \stkapriltwentysix~ -- \stkaugustseventeen ~ period, the STK temperature has been nearly constant, thus the noise has also been stable.
As the number of noisy channels is very limited after \stkaugustseventeen, it is thus possible to estimate the noise dependence with temperature as shown in Fig.~\ref{fig:avgnoisevstempfit.pdf}. A linear fit shows that the temperature dependence of the noise is of 0.008 ADC per Celsius degree.

\subsection{VA calibration}

Each STK ladder is equipped with 6 VA140 chips produced by IDEAS~\cite{cite:IDEAS}, dedicated to the signal shaping and amplification. Therefore, the STK has a total of 1152 chips that might have slightly different response gain. For this reason the on-orbit calibration is necessary to ensure an accurate signal response, as also shown in~\cite{cite:VERTEX2016}.

\begin{figure}[!htb]
	\centering
	\includegraphics[scale=0.6]{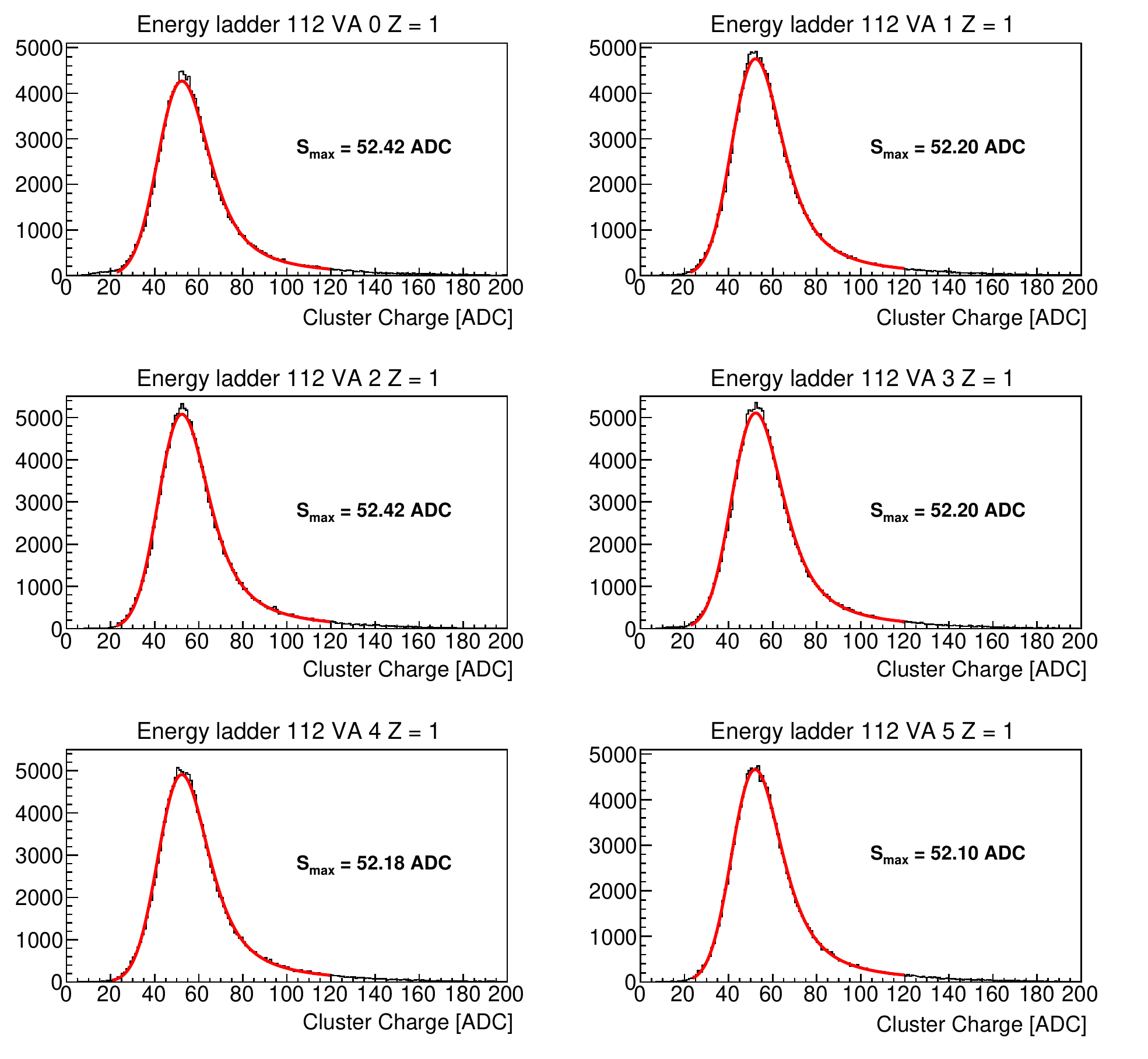}	
	\caption{Signal distribution for 6 VA140 chips belonging to the same ladder, after the VA calibration.  Each distribution is fitted with a Landau convoluted with a Gaussian function (red line).  The most probable value of the fit function, $\rm{S_{max}}$, is shown for each VA. }
	\label{fig:VACalib}
\end{figure}

\begin{figure}[htbp]
     \begin{minipage}{0.48\textwidth}
      \centering
	\includegraphics[width=0.9\columnwidth]{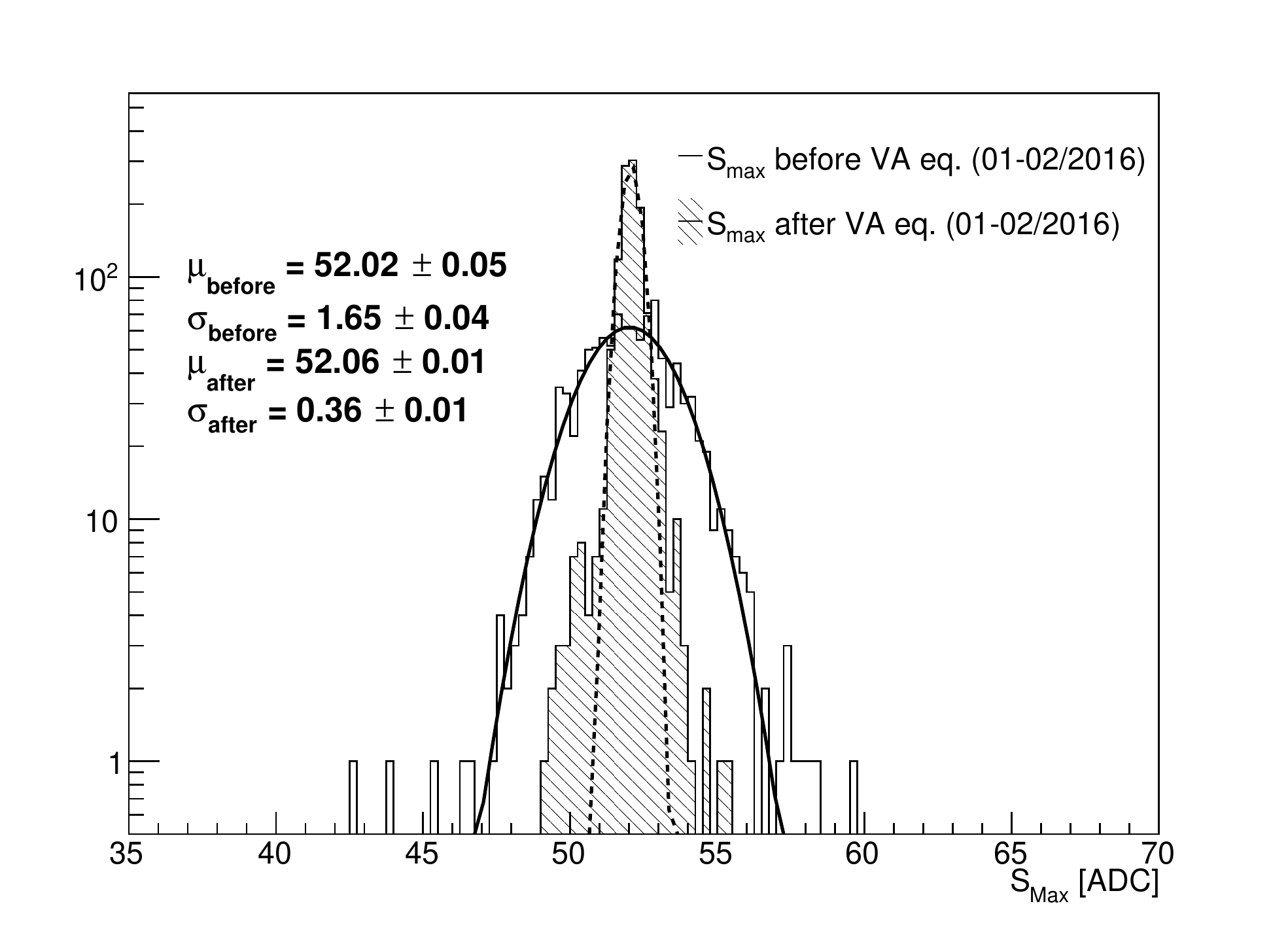}	
     \end{minipage}\hfill
     \begin{minipage}{0.48\textwidth}
      \centering
	\includegraphics[width=0.9\columnwidth]{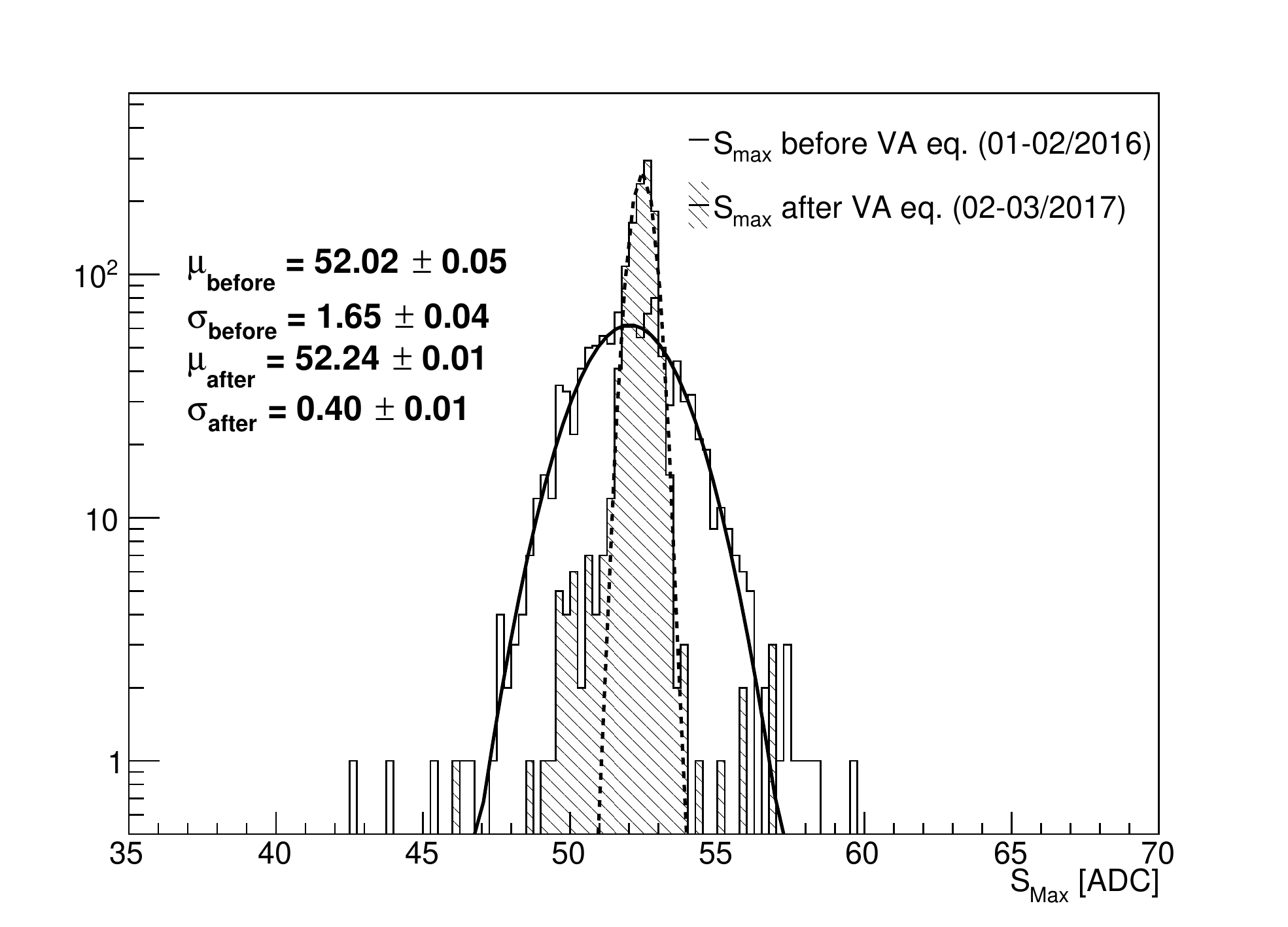}	
       \end{minipage}
	\caption{$\rm{S_{max}}$ response before (empty distribution) and after (filled distribution) the VA chip gain calibration for two different time intervals: January-February 2016 (left) and February-March 2017 (right).}
	\label{fig:VAMean}
\end{figure}

The calibration is performed using the proton candidates selected in the orbit data of two months (January and February 2016). As an example, in Fig.~\ref{fig:VACalib} we show the signal distribution for 6 VA140 chips belonging to the same ladder, where the signal is corrected for the particle path length in the silicon. 
Each distribution is fitted with a Landau convoluted with a Gaussian function, where the maximum probable value of the fit, $\rm{S_{max}}$, corresponds to an optimal chip response. The distribution of $\rm{S_{max}}$  for all VA chips of the STK is shown in Fig.~\ref{fig:VAMean}.

Before the calibration procedure is applied, the signal response extends from 44 to 60 ADC counts (non-filled distribution in Figure~\ref{fig:VAMean}). Each chip is therefore calibrated to the mean value of this distribution. The chip signal response after applying the correction is shown in Fig.~\ref{fig:VAMean} for two different time periods: January-February 2016 and February-March 2017. As a result of the calibration, the spread  of  $\rm{S_{max}}$ for different VA chips is less than 0.5 ADC counts,  ensuring therefore a uniform signal response of the STK. The average $\rm{S_{max}}$ remains stable on-orbit, the total change with time is less than 1 ADC counts after more than one year of operation.

\section{BGO calorimeter calibration}

The BGO calorimeter is designed to measure the energy of incident particles, to determine the direction of particles, and to separate electrons from protons \cite{Chang2014,Chang2017,DAMPE2017}.
The BGO calorimeter contains 14 layers of BGO crystals with a total radiation length of 32 and a nuclear interaction length of 1.6. Each layer is composed of 22 BGO crystal bars, which are read out by PMTs on two sides. Each PMT outputs signals from 3 different dynodes, i.e. Dy8 (high gain channel), Dy5 (medium gain channel) and Dy2 (low gain channel). The crystal bars are supported by a carbon fiber structure in a hodoscopic arrangement to measure the energy deposition and profiles of hadronic and electromagnetic showers developed in the calorimeter. The shower topology is especially crucial for the electron/proton separation \cite{Chang2008b,DAMPE2017}.

A detailed description of the calorimeter has been published elsewhere \cite{Chang2017}. Here we describe the on-orbit calibration of the BGO calorimeter, including: (A) the pedestal calibration, (B) the MIP response calibration, (C) the PMT dynode ratio calibration, (D) the light attenuation calibration, and (E) the trigger threshold.

\subsection{Pedestal calibration}
\begin{figure} \centering
{
\includegraphics[width=0.32\columnwidth]{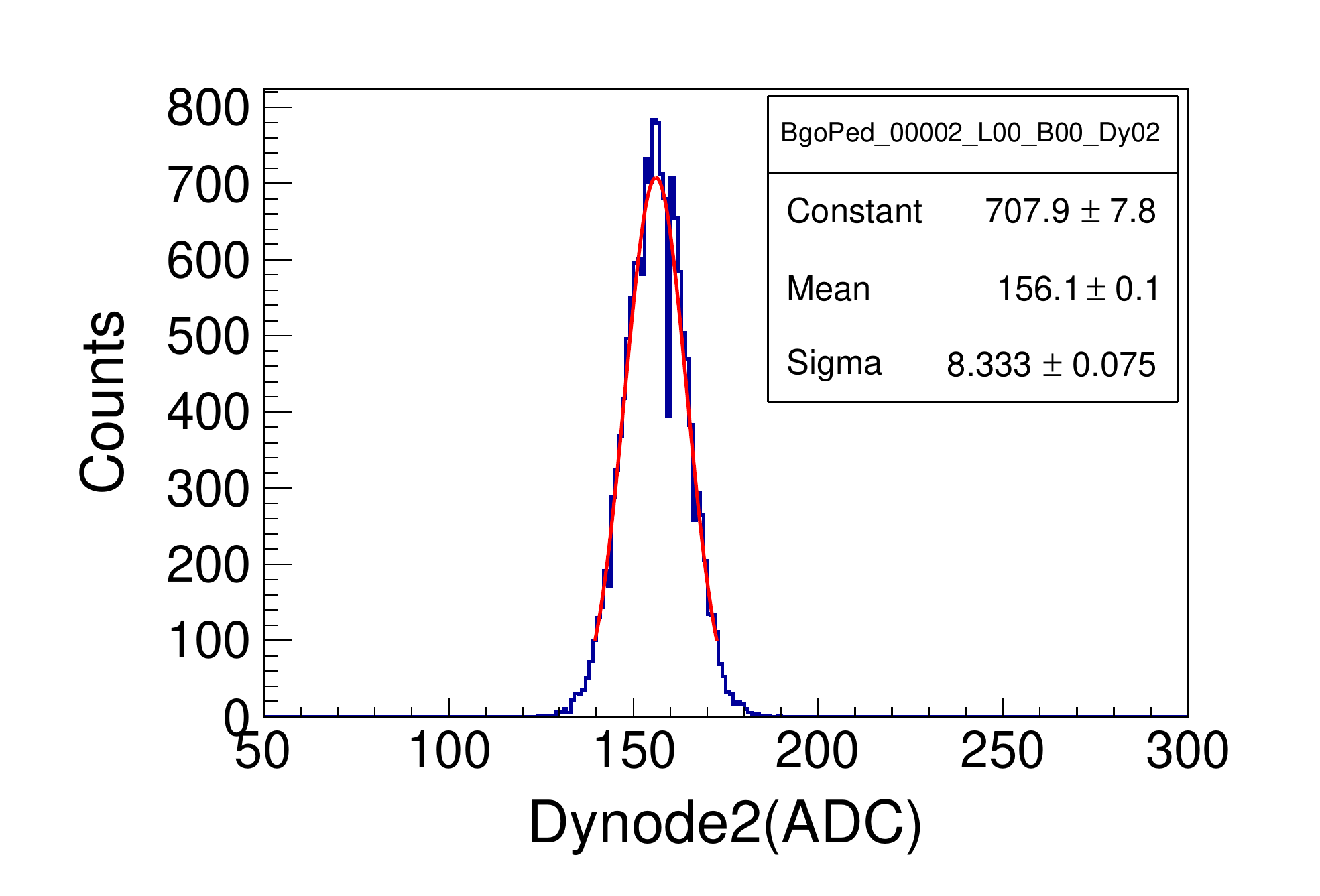}
\includegraphics[width=0.32\columnwidth]{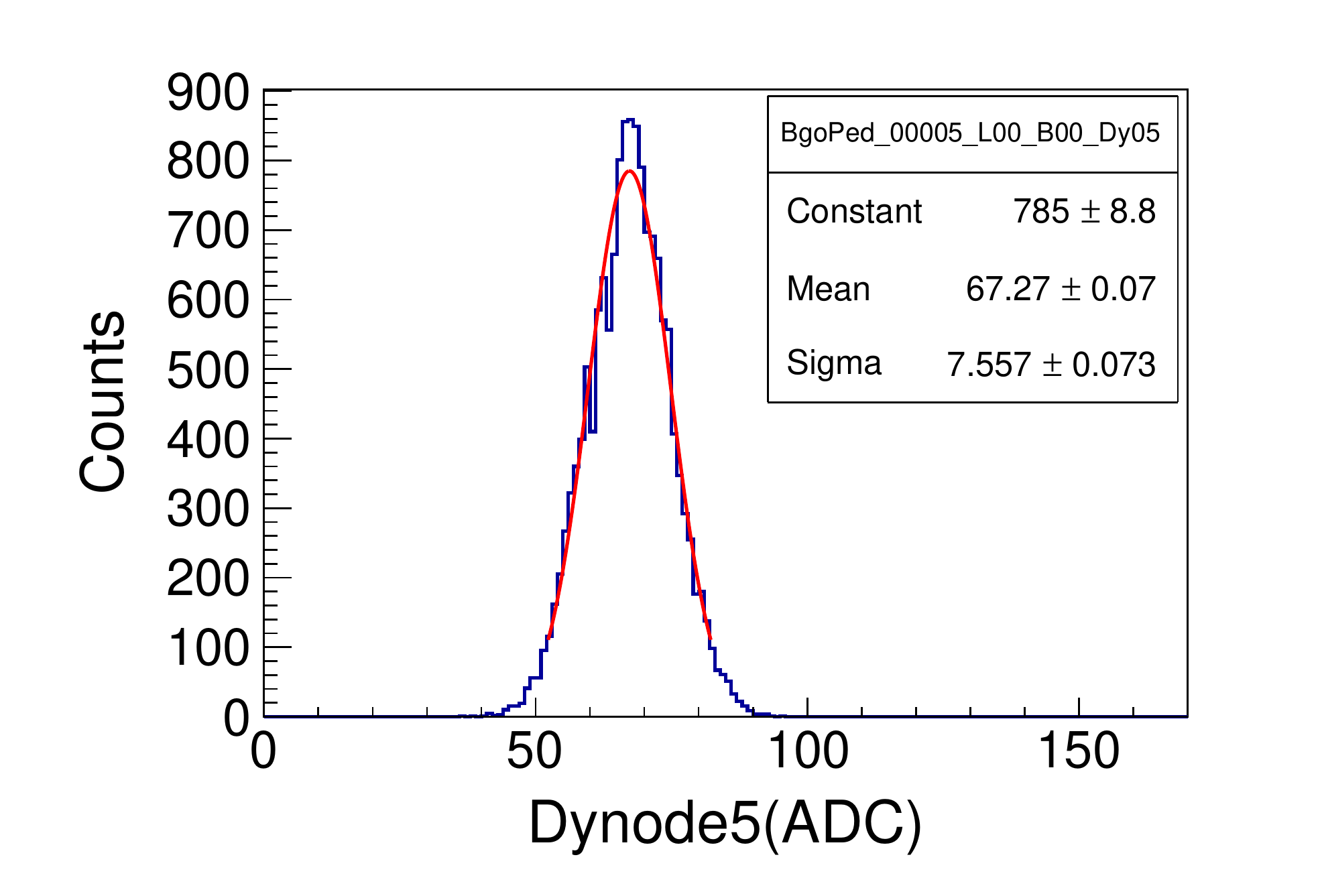}
\includegraphics[width=0.32\columnwidth]{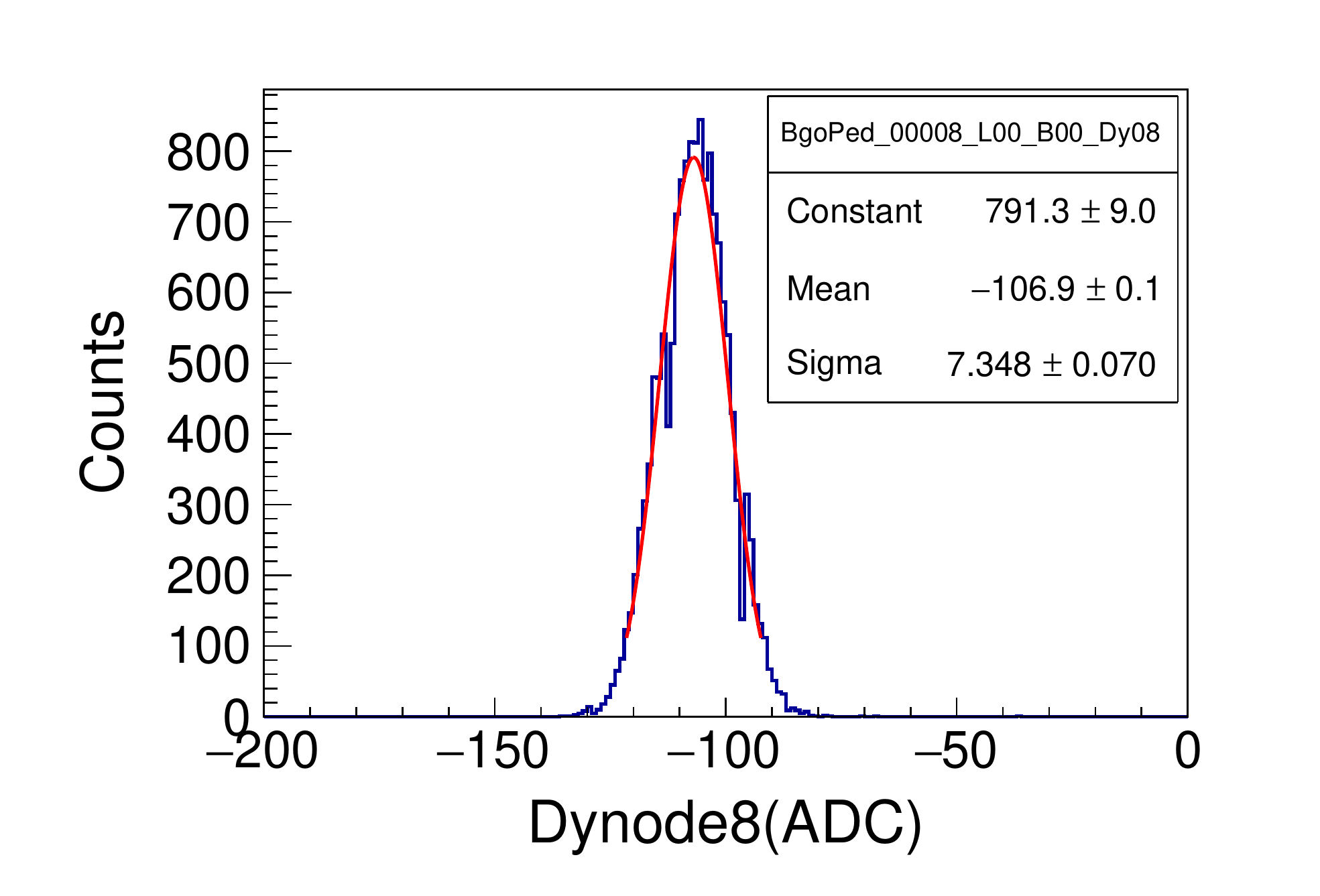}
}\caption{The pedestal distributions of Dy2 (left), Dy5 (middle), Dy8 (right) of a PMT. The Gaussian fits to the distribution are shown as red curves.}
\label{fig:BGO-dy258}
\end{figure}

On-orbit pedestals are calibrated using periodic triggers. Fig.~\ref{fig:BGO-dy258} shows the typical pedestal distributions of the three different dynodes of a PMT. The pedestal distribution peaks between $-400$ and $400$ ADC units. The pedestal widths, which represent the noise level, are about 8 ADC units, consistent with the ground test results \cite{ZhangZY2016}.

\begin{figure} \centering
\includegraphics[width=0.45\columnwidth]{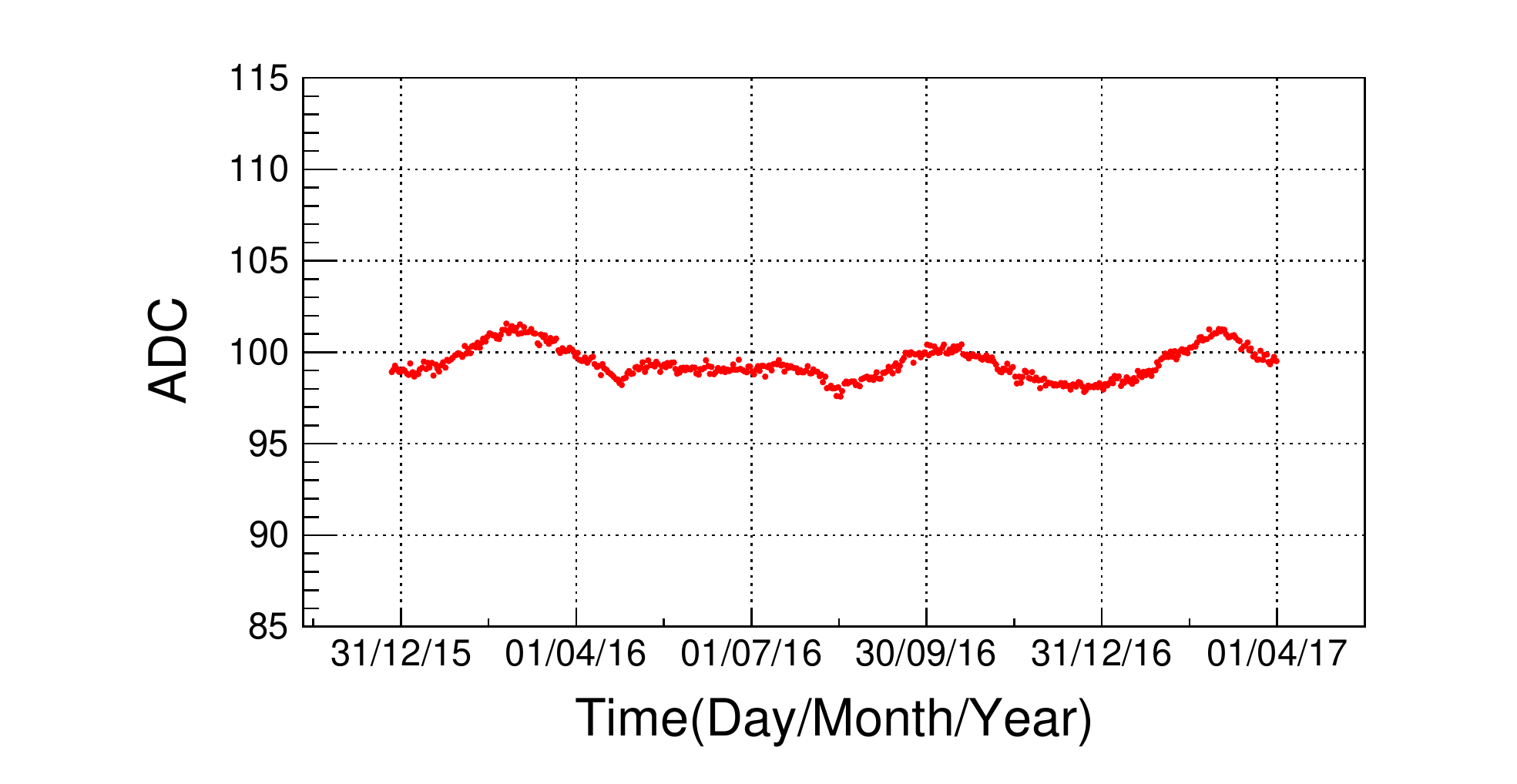}
\includegraphics[width=0.45\columnwidth]{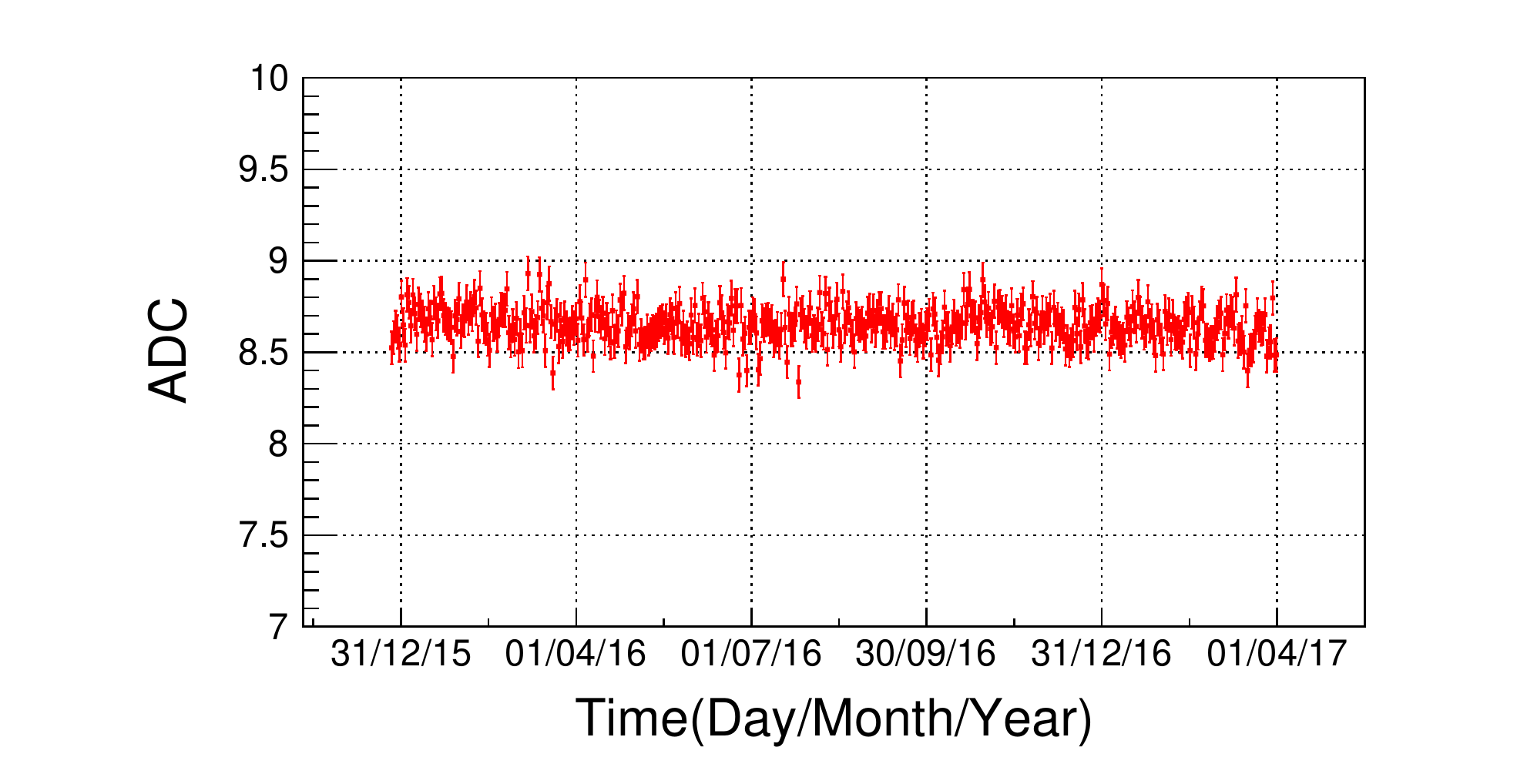}
\caption{The pedestal mean value (left) and width (right) as functions of time.}
\label{fig:BGO-Ped-stablity}
\end{figure}

The pedestal values are continuously monitored during the on-orbit operation. The results are very stable. Fig.~\ref{fig:BGO-Ped-stablity} shows the stability of the on-orbit pedestal for about 15 months. The drift of the mean value is very small (within $\pm2$ ADC, see Fig.~\ref{fig:BGO-Ped-stablity}(a)) and the the sigma value is also very stable (see Fig.~\ref{fig:BGO-Ped-stablity}(b)). The small drift of the mean value is caused by the temperature change of the BGO calorimeter. The drift of the pedestal is less than 1 ADC unit per degree.

\subsection{MIP calibration}

\begin{figure} \centering
\includegraphics[width=0.8\textwidth]{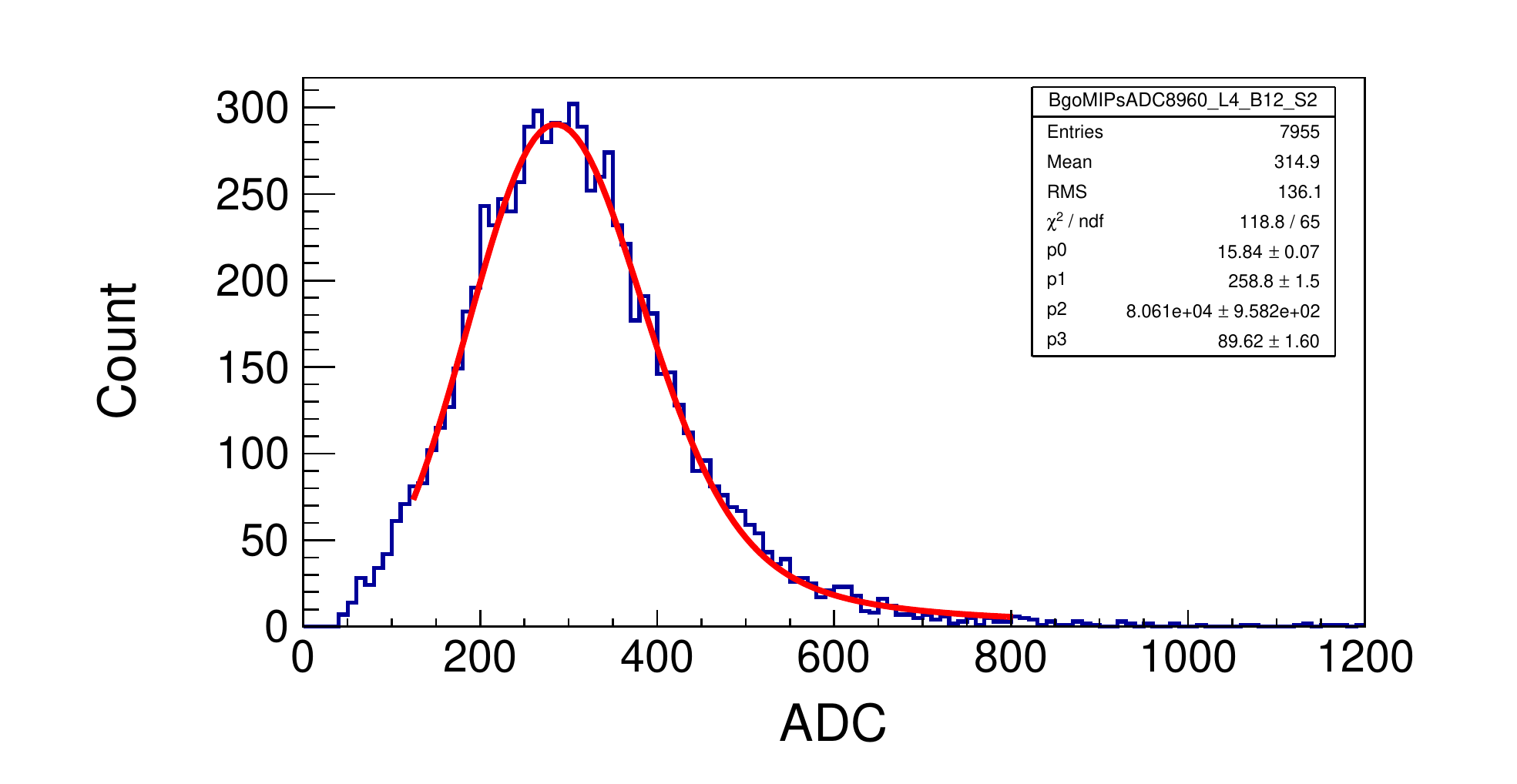}
\caption{The MIP spectrum of a BGO bar.}
\label{fig:BGO-MIPs_bgo}
\end{figure}

The calibration of individual crystal energy scale involves the determination of the parameters of a transfer function between the energy deposited in the crystal and the signal output in digital counts. There are some differences between the on-orbit and the ground calibration. The energy scale was calibrated using muon MIPs on the ground, while it was calibrated using proton MIPs on-orbit. About $80-90\%$ of the incident protons will interact within the BGO calorimeter via hadron-nucleon interaction. Only a small fraction of proton events are MIPs which can be used to calibrate the calorimeter. Fig.~\ref{fig:BGO-MIPs_bgo} shows the MIP spectrum of a BGO bar. In order to determine the peak position, a Gaussian convolved Landau function is used to fit the spectrum. The best-fit MPV is obtained and used for the absolute energy calibration.

\begin{figure} \centering
\includegraphics[width=0.8\textwidth]{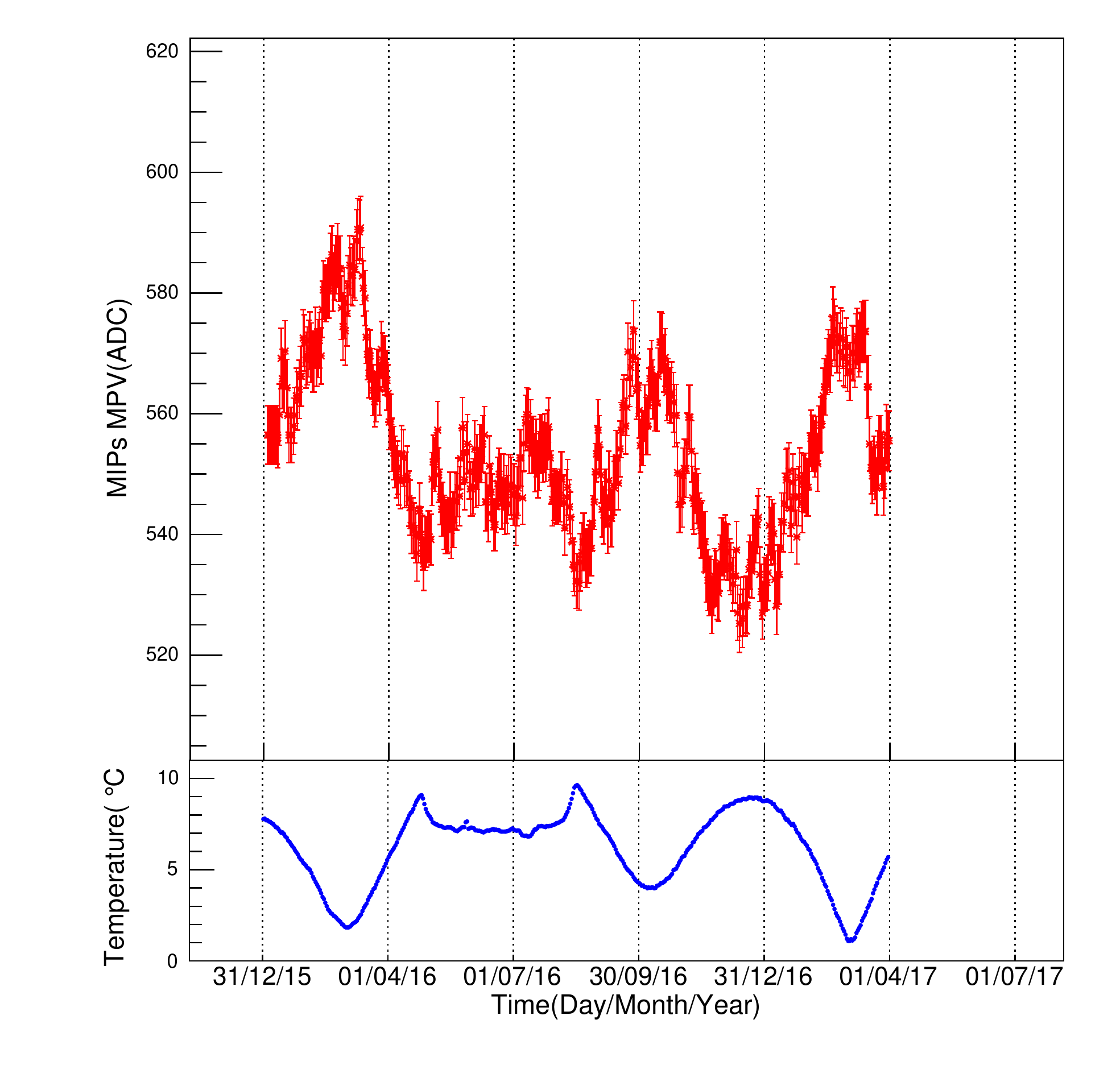}
\caption{Variations of the MIP MPV value (upper panel) and the BGO temperature (lower panel) with time.}
\label{fig:BGO-MIPs_temp}
\end{figure}

\begin{figure} \centering
\includegraphics[width=0.8\textwidth]{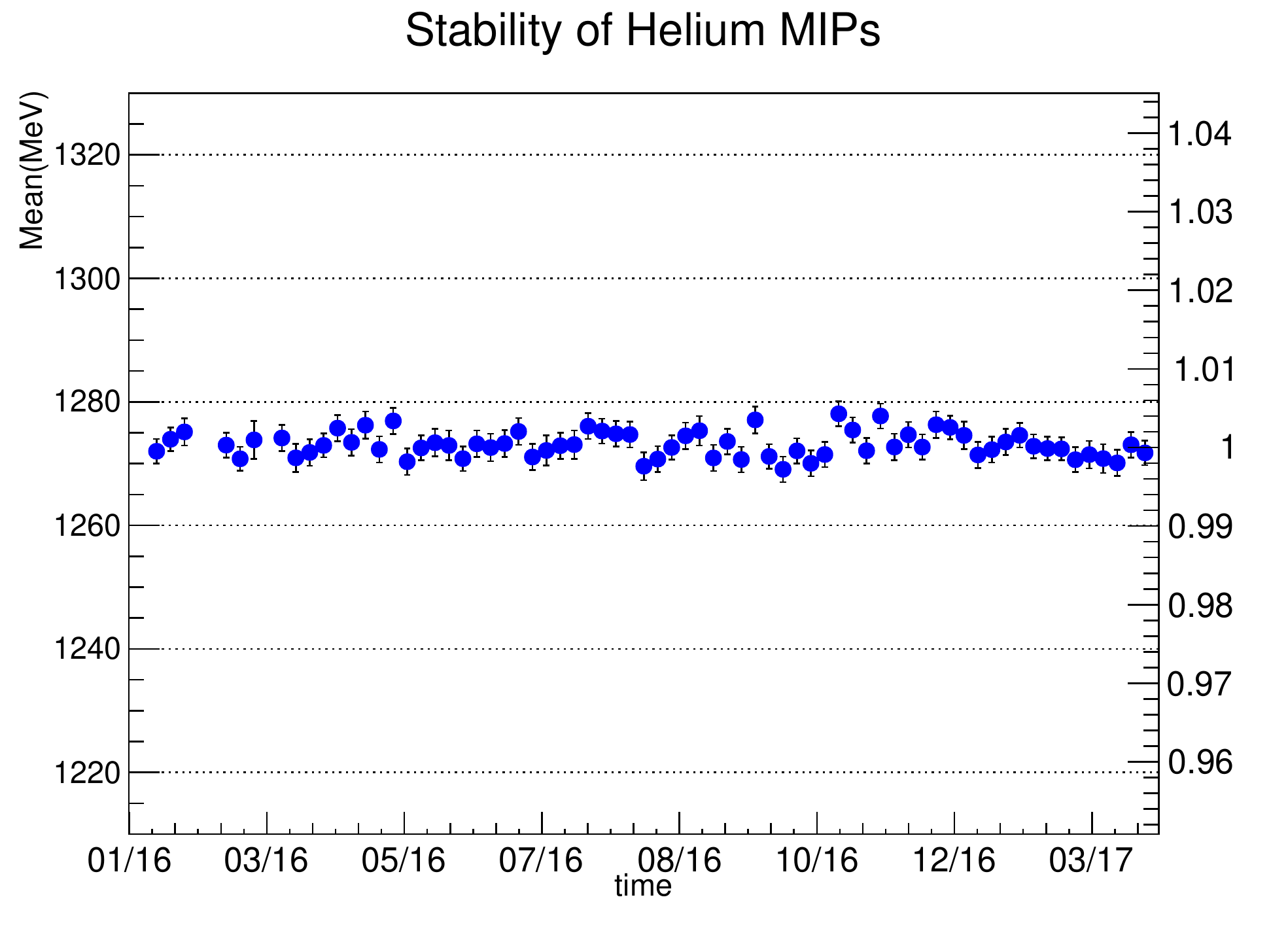}
\caption{Evolution of temperature corrected total energy deposition of helium MIPs in the BGO calorimeter.}
\label{fig:BGO-Helium}
\end{figure}

The MIP calibration trigger is only allowed near the equator (for latitudes within $\pm 20^{\circ}$) to suppress the influence of low energy (i.e., non-relativistic) events. This is because the geomagnetic cutoff is higher than 10 GV near the equator, and the flux of non-relativistic protons is very low. Fig.~\ref{fig:BGO-MIPs_temp} shows the daily variation of the MPV values of the MIP spectra and the corresponding temperature variation. Due to the temperature effect on the light yield of BGO crystals, the MPV value and the temperature are roughly anti-correlated. To correct the temperature effect, the MIP calibration has been done in every orbit. We have also carried out the calibration using Helium MIP events. Fig.~\ref{fig:BGO-Helium} shows the stability of total energy deposition in the BGO calorimeter of helium MIP events. The variation is less than 1\% after the temperature correction.

\subsection{PMT dynode ratio calibration}

\begin{figure} \centering
\includegraphics[width=0.45\columnwidth]{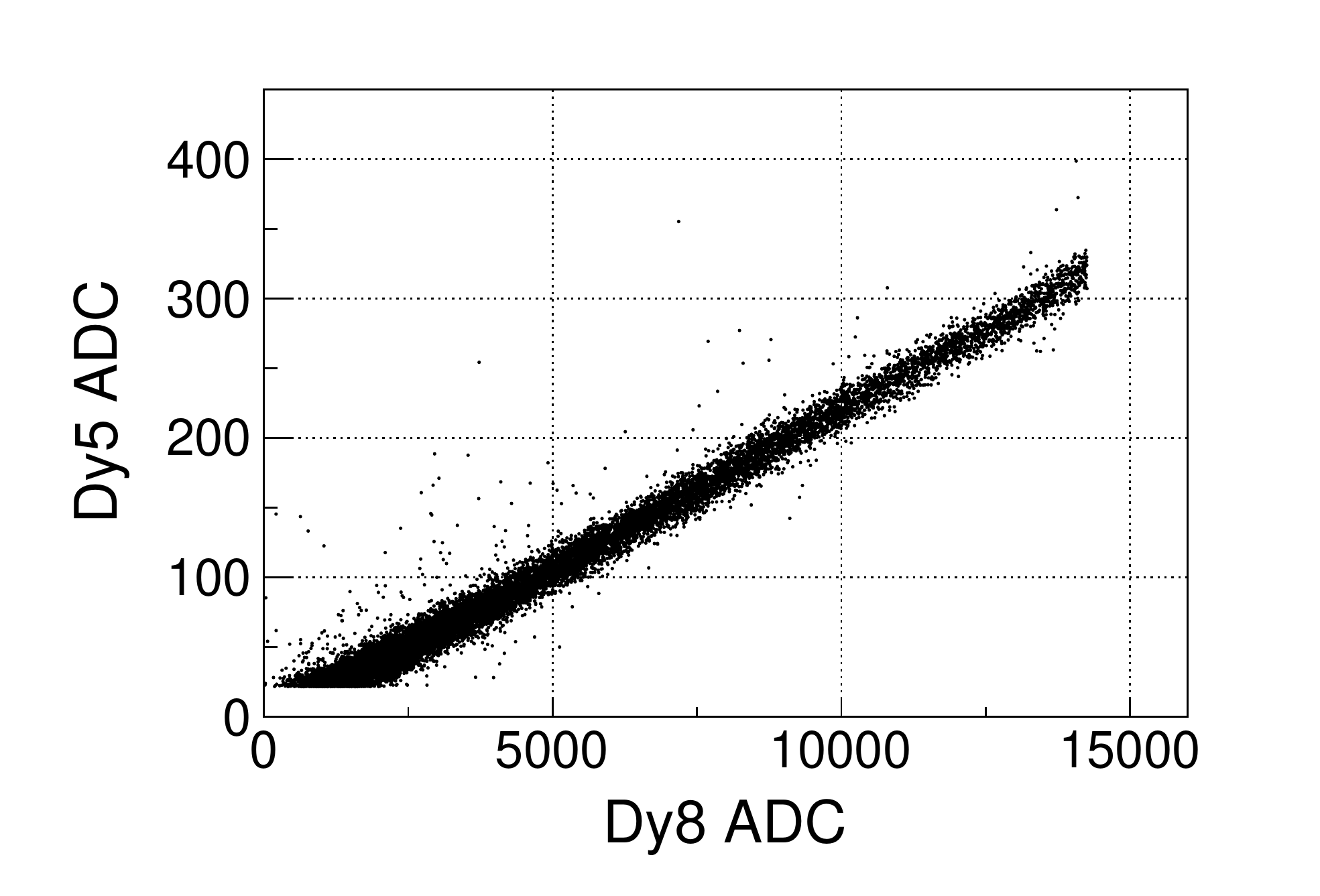}
\includegraphics[width=0.45\columnwidth]{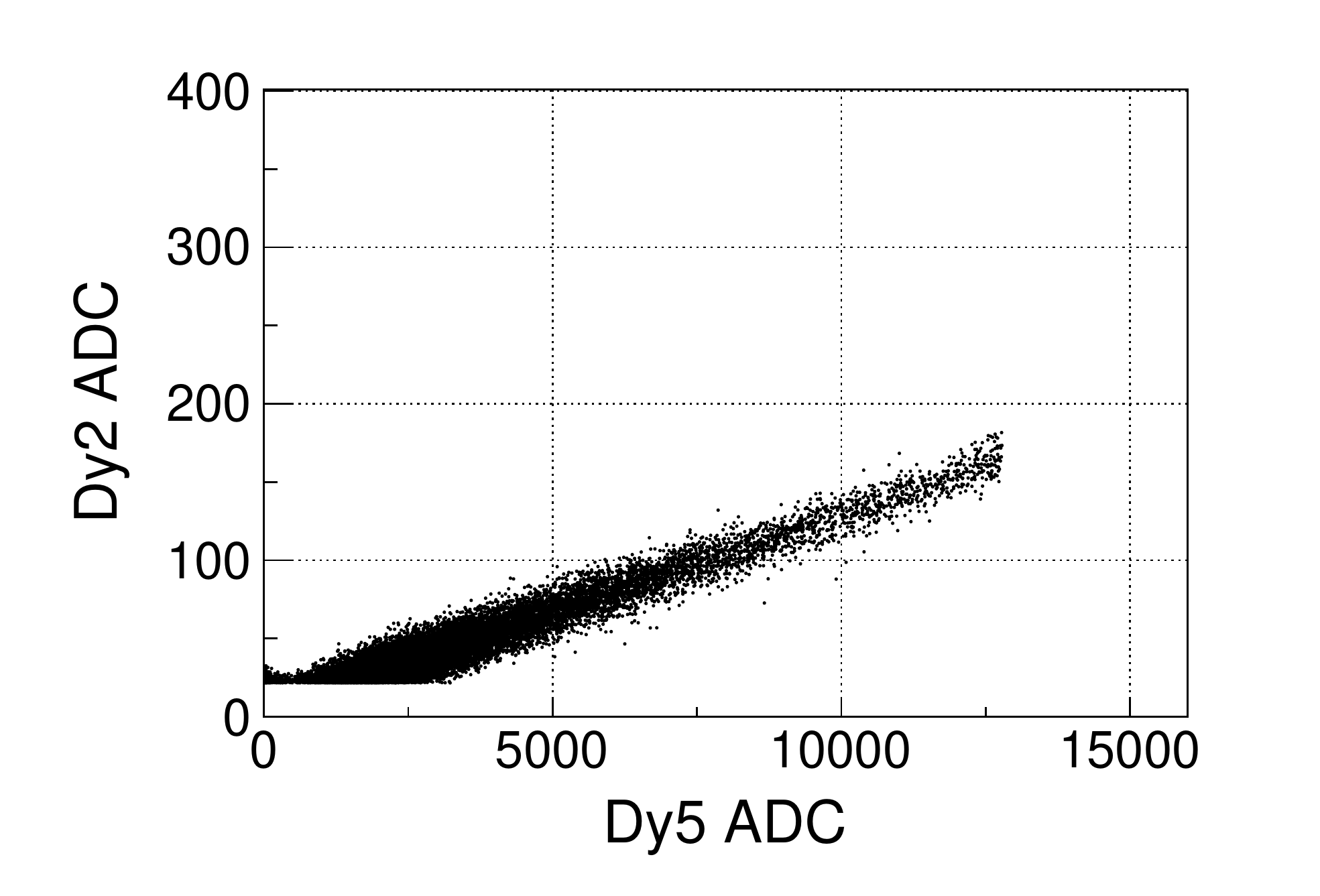}
\caption{The Dynode ratios Dy8/Dy5 (left panel) and Dy5/Dy2 (right panel) of a PMT.}
\label{fig:BGO-dy258Ratio}
\end{figure}

\begin{figure} \centering
\includegraphics[width=0.45\columnwidth]{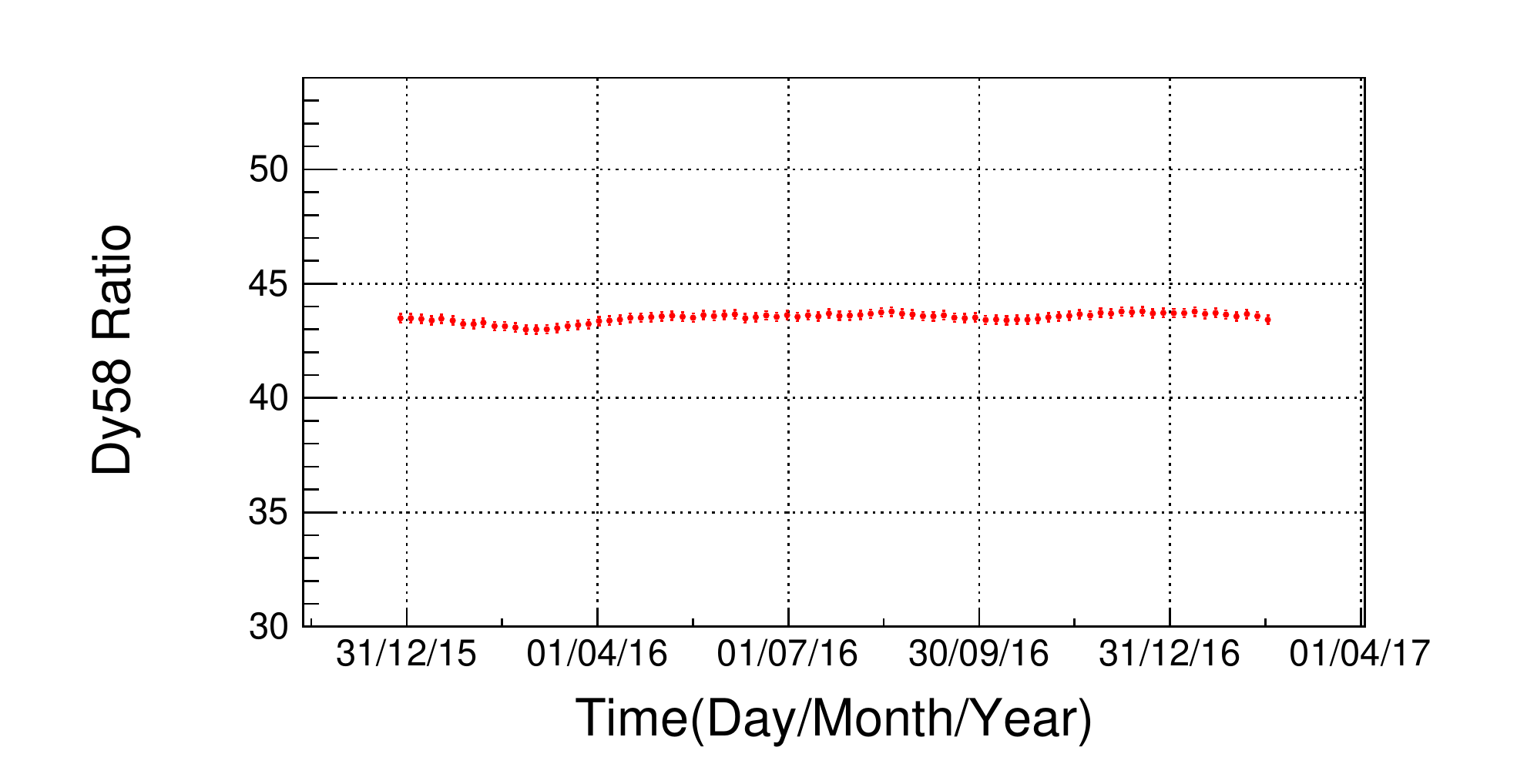}
\includegraphics[width=0.45\columnwidth]{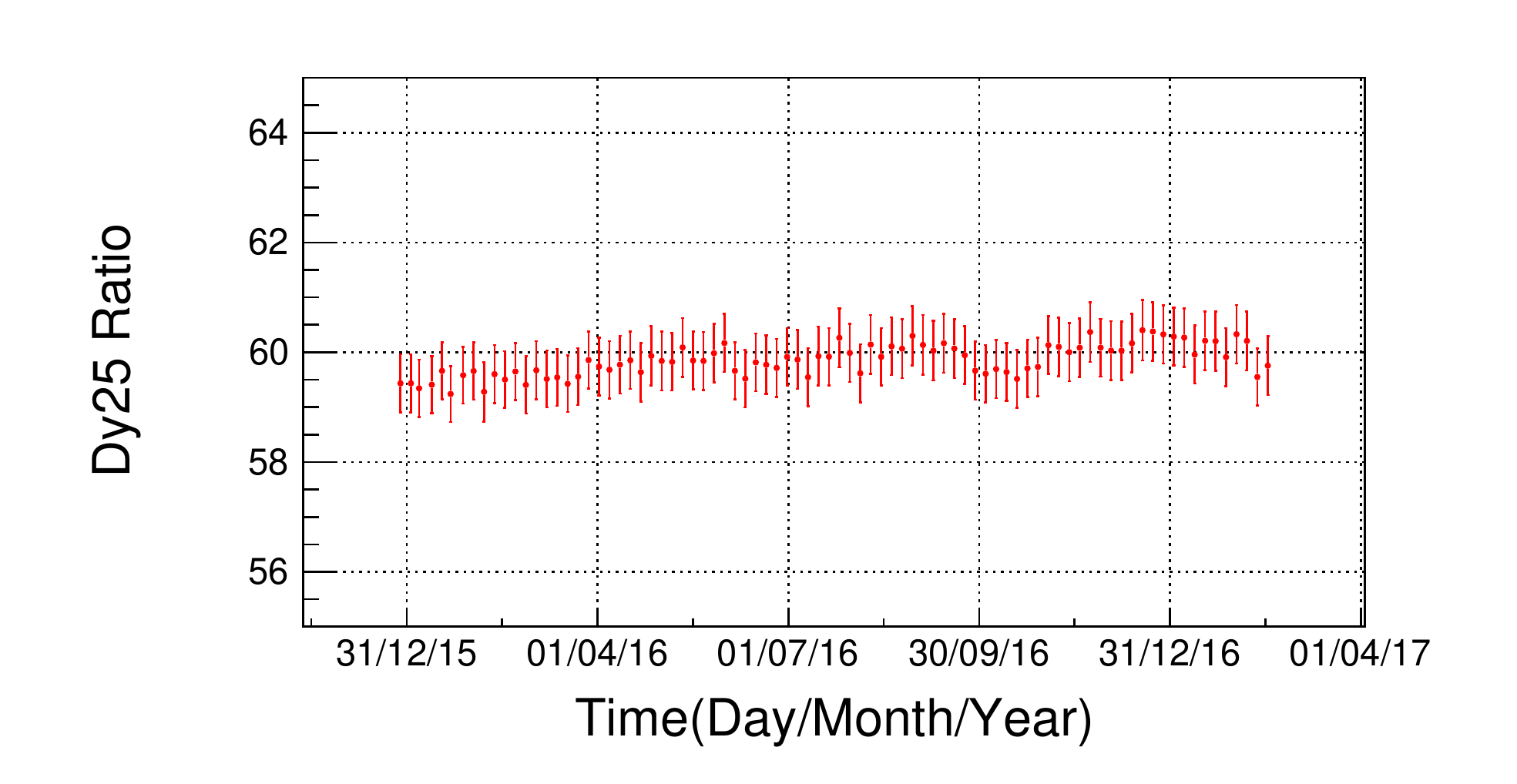}
\caption{Variations of dynode ratios Dy8/Dy5 (left panel) and Dy5/Dy2 (right panel) in time.}
\label{fig:BGO-dy258time}
\end{figure}

The DAMPE detector is designed to cover a wide energy range from $\sim5$ GeV to $\sim10$ TeV for electrons and $\gamma$-rays. For such a purpose, the sensitive element of the BGO calorimeter should cover the dynamic range from 10 MeV ($\sim 0.5$ MIPs) to 2 TeV ($\sim 10^{5}$ MIPs). To reach such a wide dynamic range, a multi-dynode readout method is adopted \cite{ZhangZY2015}. In order to calibrate the relative dynode ratio, { both electromagnetic and hadronic shower events with high energy deposition in the BGO crystal bar were used} \cite{DongJN2017}. Fig.~\ref{fig:BGO-dy258Ratio} shows the linear correlation of different dynodes. The ratio is $\sim44$ for Dy8/Dy5 and $\sim60$ for Dy5/Dy2. In order to evaluate the stability of the PMT performance for a long time, the ratios are computed day by day. The results are shown in Fig.~\ref{fig:BGO-dy258time}. During the first 15-month's operation of DAMPE in space, the ratios of Dy8/Dy5 and Dy5/Dy2 are very stable.

\subsection{Light attenuation calibration}

\begin{figure} \centering
\includegraphics[width=0.8\textwidth]{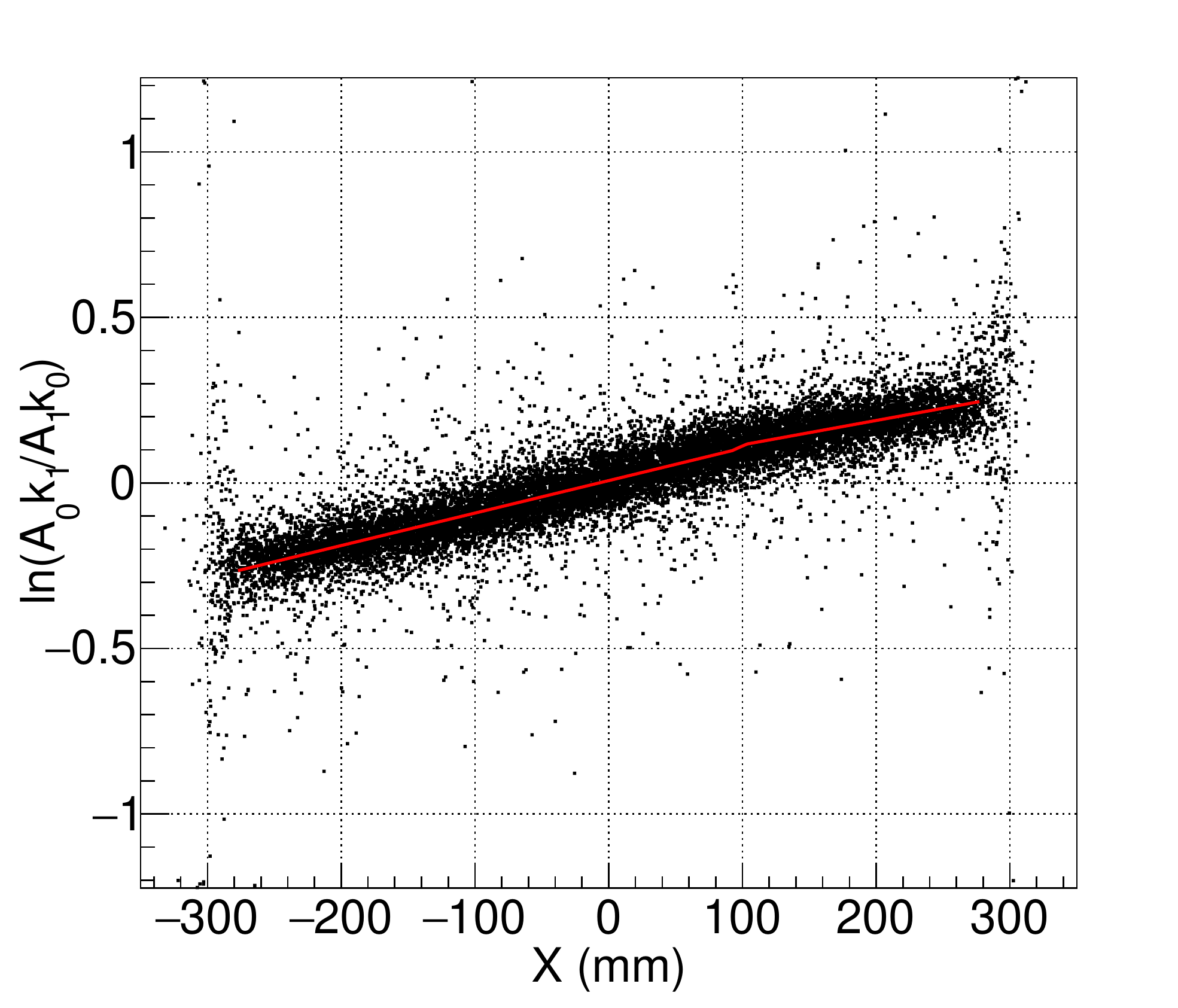}
\caption{The light asymmetry measured in a typical BGO bar using shower events.}
\label{fig:BGO-attenuationlength}
\end{figure}

In order to reliably infer the position of a hit through comparing the signals obtained from both ends of a crystal, the light attenuation length of a BGO bar should be calibrated.
Considering the small moliere radius of BGO crystal, the electron candidates are used for this calibration. An exponential function is adopted to describe the light attenuation behavior in BGO crystal. Signals from the positive and negative sides of a crystal are denoted as $A_{0}$ and $A_{1}$,
\begin{equation}
\begin{array}{cll}
A_{0} = k_{0}A_{p}e^{-x/\lambda},\\
A_{1} = k_{1}A_{p}e^{-(L-x)/\lambda},
\label{eq:BGO-1}
\end{array}
\end{equation}
{ where $A_{p}$ is the primary light yield produced in the crystal, $k_{0}$ and $k_{1}$ are the conversion coefficients between the light at the corresponding end of the crystal, and $x$ is the hit position of BGO bar that satisfies the relation}
\begin{equation}
\begin{array}{cll}
x = \frac{L-\ln\frac{A_{0}k_{1}}{A_{1}k_{0}}\lambda}{2},
\end{array}
\label{eq:BGO-2}
\end{equation}
where $L$ is length of the BGO crystal, and $\lambda\approx1.5$ m is the attenuation length determined by fitting the relation between $\ln({A_{0}k_{1}}/{A_{1}k_{0}})$ and $x$, as shown in Fig.~\ref{fig:BGO-attenuationlength}.


\subsection{Trigger threshold}
\begin{figure} \centering
\includegraphics[width=0.8\textwidth]{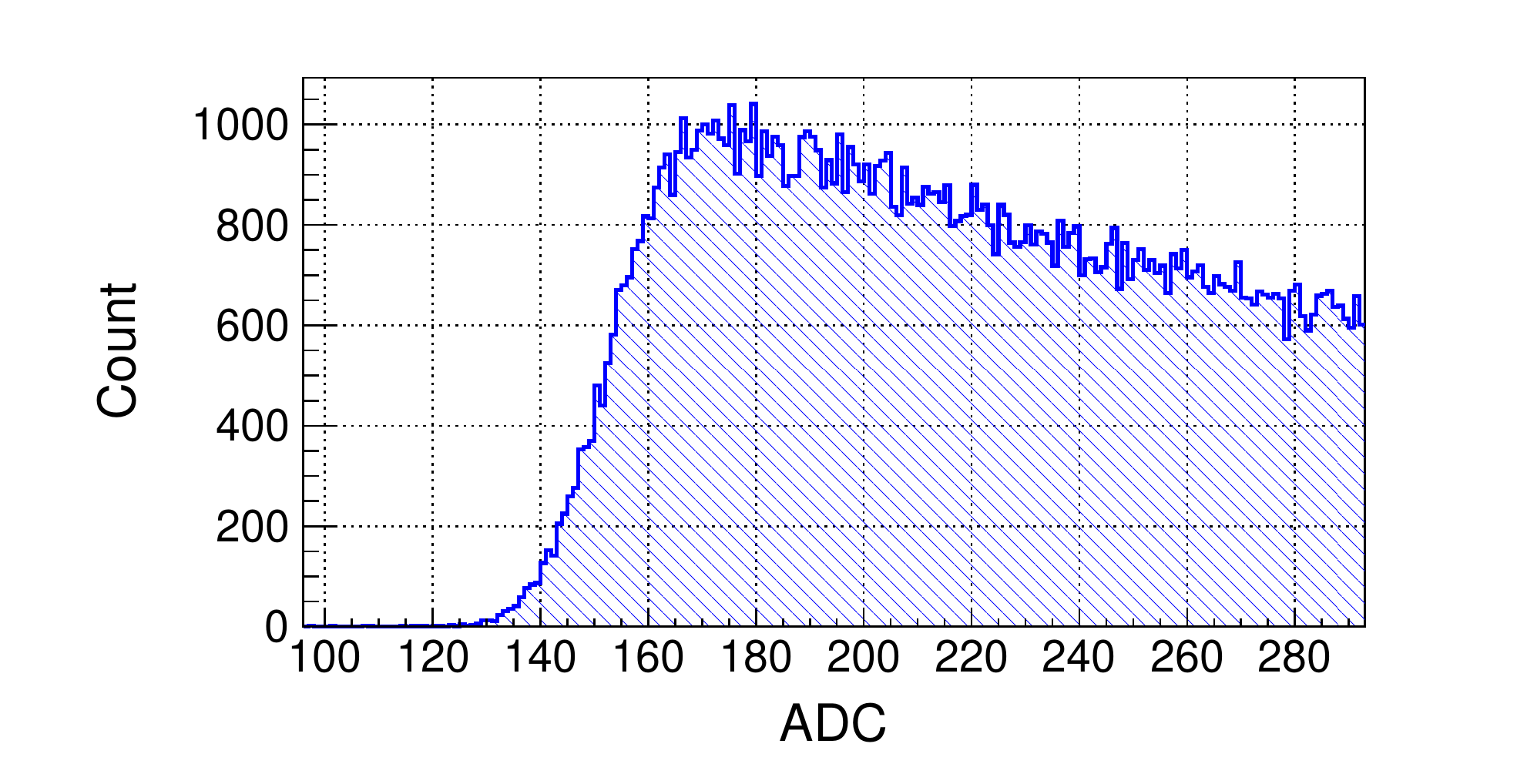}
\caption{The signals from one channel after the threshold setting.}
\label{fig:BGO-threshold}
\end{figure}

The BGO calorimeter also provides hit signals for the trigger system. The calorimeter front end electronics chip VATA160 is composed of the VA part and the TA part, both with 32 channels. The VA part is exactly the same as the VA160 chip, and the TA part is an advanced version of TA32CG ASIC to generate fast trigger \cite{FengCQ2014}.
The integrated pulse from the charge sensitive pre-amplifier (of the VA part) is firstly shaped into a fast narrow pulse and then discriminated by a comparator, to generate a digital signal. All the 32 outputs of the TA part are internally ORed together, and the output signal is shared by all input channels.
Different logics and thresholds are set for various physical goals. The calibration of the trigger threshold provides energy threshold of each bar for DAMPE's precise simulation to trigger efficiency. { Due to the inconsistency of the individual detection units,} the thresholds should be calibrated channel by channel.
Fig.~\ref{fig:BGO-threshold} shows the fired (over threshold) ADC distribution from one typical channel after the threshold cut. In principle there should be a sharp ``cut off" caused by the channel's threshold. Because of the noise superposing on the input signal at the comparator end, the count does not jump to zero directly but drops quickly when the channel's signal approaches to the threshold. The ADC value whose counts are 50\% of the maximum counts in the distribution is selected as the channel's hit threshold in ADC unit. Combining the energy scale calibration, the ADC unit for the trigger thresholds can be converted into MeV units.

\section{NUD calibration}

\begin{figure} \centering
\includegraphics[width=0.65\textwidth]{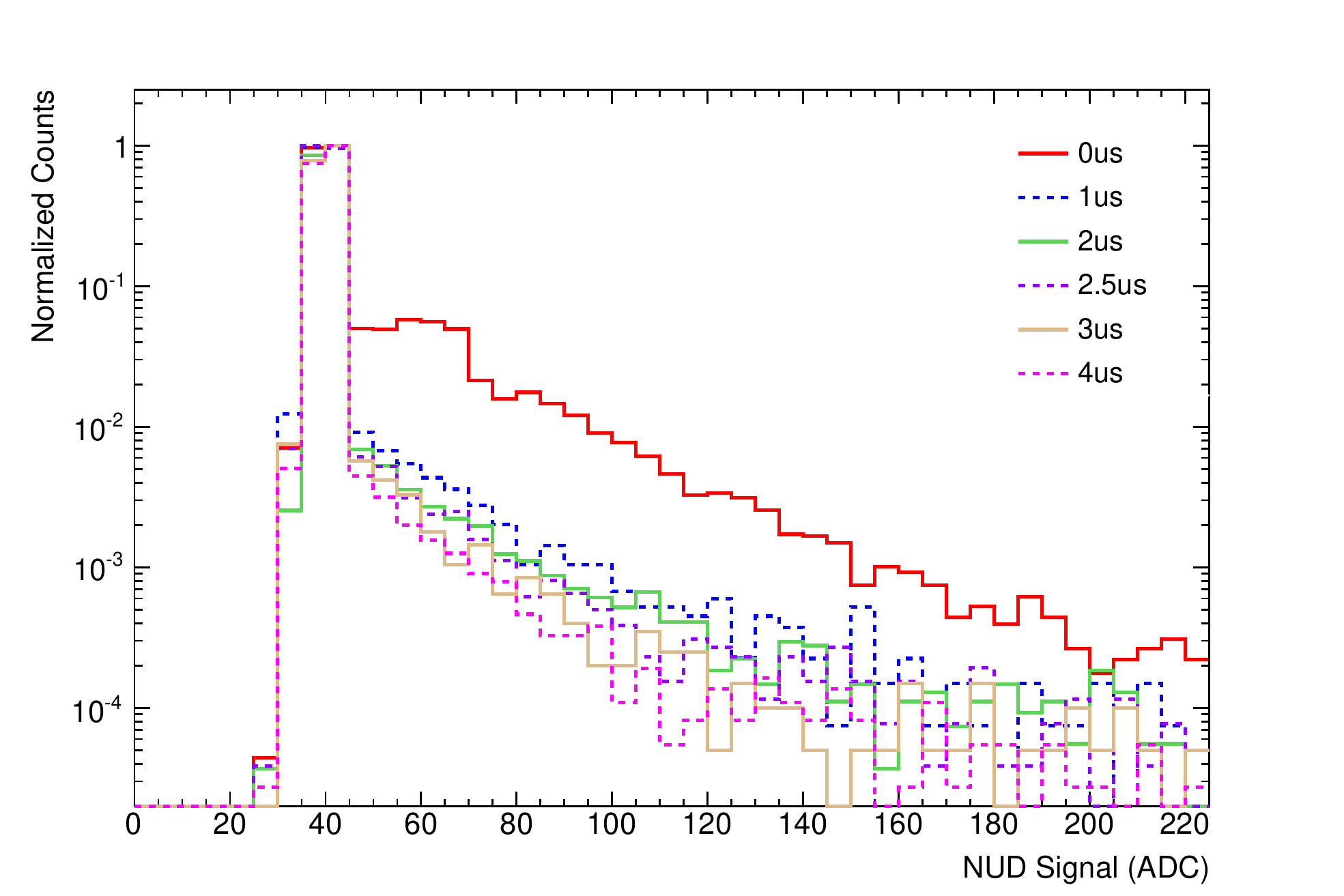}
\caption{NUD response spectrum to protons with different setups of the gate opening time after trigger.}
\label{fig:NUD-DiffTime}
\end{figure}

The NUD equipped beneath the BGO is designed to detect the signals of secondary neutrons of BGO shower tails. However, the NUD signals within $\sim2\ \mu$s after the trigger are dominated by secondary charged particles. We thus record only the signals $\sim2\ \mu$s after the trigger which mainly come from secondary neutrons. The exact delay time of the NUD integrator's gate in respect to the trigger needs to be carefully tuned \cite{Chang2017}. Fig.~\ref{fig:NUD-DiffTime} shows the normalized histogram of NUD counts versus delay time from 0 $\mu$s to 4 $\mu$s. For a delay time of less than 2 $\mu$s, the signals are the mixture of both neutrons and secondary charged particles as indicated by the relatively flat tail of high-ADC signals. With increasing delay time, the counts are highly suppressed. For delay time of 2, 2.5, 3, and 4 $\mu$s, there is no significant difference in the profiles and the signals should be dominated by secondary neutrons. The gate is thus configured to a delay time of 2.5 $\mu$s.

\begin{figure} \centering
\includegraphics[width=0.65\textwidth]{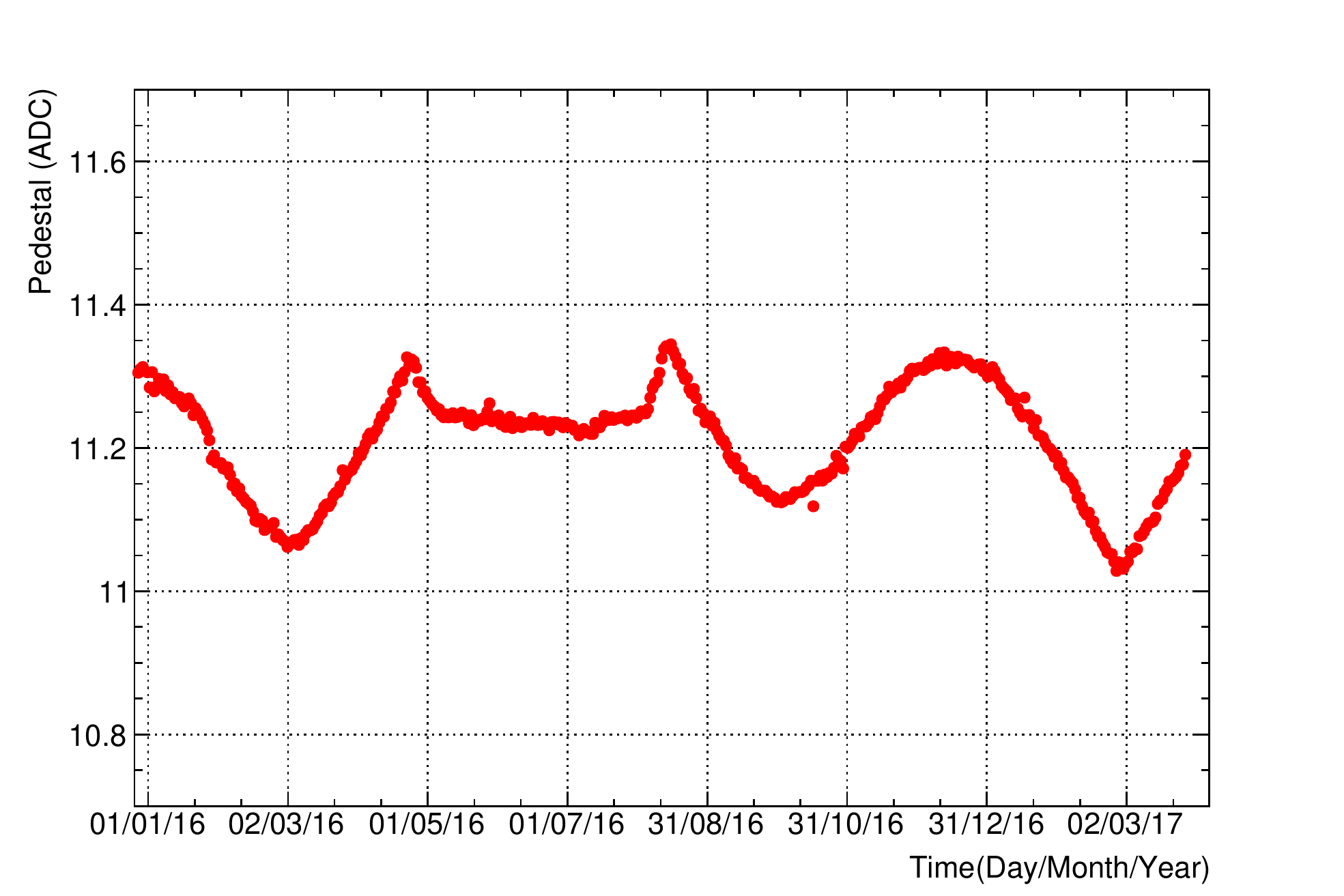}
\caption{ The pedestal variation of a typical NUD channel in time.}
\label{fig:NUD-Pedestal}
\end{figure}

{ The pedestal of NUD is also calibrated using periodic trigger events every day, which is the same as BGO and PSD. Fig.~\ref{fig:NUD-Pedestal} shows the stability of a typical NUD pedestal for about 16 months. The on-orbit NUD pedestal drifts by less than 0.4 ADC, indicating a very stable status of NUD during the whole period of operation.}


\section{SAA perimeter evaluation}

Due to non-concentricity of the Earth and its dipole-like magnetic field, the Van Allen radiation belt dips down to $\sim200$ km in the so-called SAA region. The fluxes of trapped protons and electrons in the radiation belt are two orders of magnitude higher than in other regions \cite{Casolino2008,Abdo2012Calibriation}.
DAMPE is operating in a sun-synchronous orbit at an altitude of $ \sim500$ km with an inclination of $\sim97^\circ$. As a result, it travels through the SAA region about 7 times per day. The typical energy of trapped protons in the SAA is extremely low relative to DAMPE's sensitive energy range. Most of them are blocked by the shield of the satellite. Simulations and ground beam tests show that the electronics of DAMPE survives under such an environment of high-energy particles without obvious performance degradation and lifetime shortening \cite{GaoSS2014b,Zhang:2016hth}. Therefore, DAMPE maintains its normal operation in the SAA. However, the intensive particle hit rate within the SAA region may induce severe pile-up effects. The data acquired in the SAA region cannot be used for many science analyses. With the recorded data of DAMPE we develop a method to precisely determine the perimeter of the SAA region.

\begin{figure}[!htb]
\includegraphics[scale=0.8]{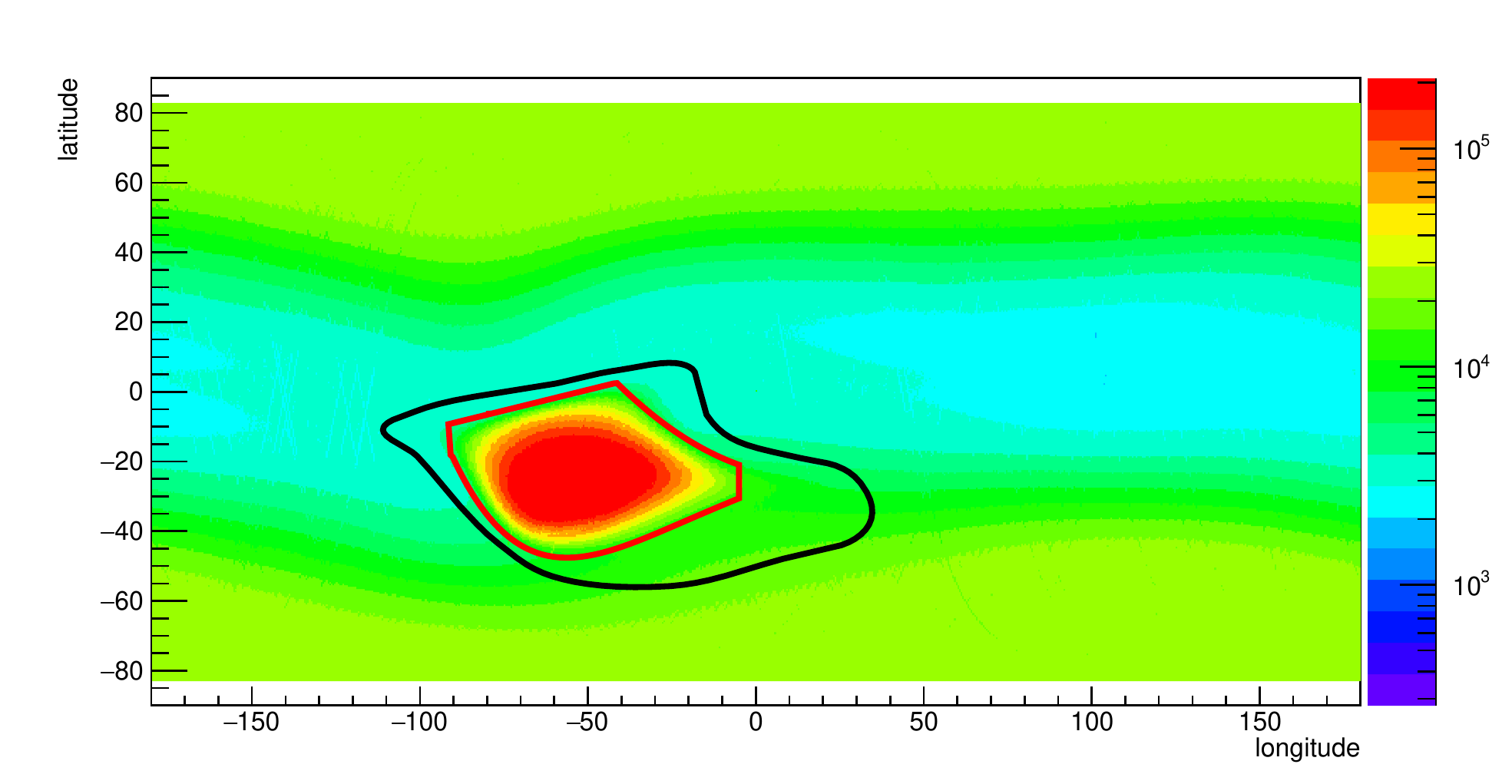}
\caption{The trigger rate at different geographic longitudes and latitudes. The red solid curve outlines the SAA boundary fitted by the data from Jan. 1 2016 to Jan. 1 2017. The black solid line is the contour of about 1 ${\rm cm^{-2}~s^{-1}}$ for trapped proton flux above 10 MeV predicted by the AP-9 \cite{Ginet2013}.} \label{fig:SAA-1}
\end{figure}

The SAA was already measured by many experiments (e.g.,\cite{Casolino2008,Abdo2012Calibriation,Ackermann2012}). The space weather forecast \cite{Ginet2013} has already defined a general map of the SAA region (see Fig.~\ref{fig:SAA-1}). It is not suitable for DAMPE because the SAA exclusive criteria of DAMPE, determined by the affection level of electronics, is different from other experiments. Shortly after the successful operation of DAMPE, we found that the hit rate given by the top BGO layers with a low trigger threshold is a direct indicator of the particle flux in space, thereby providing a good and simple way to identify the SAA region. This trigger rate is recorded every 4 seconds and transmitted to the ground as house keeping data. The map of this trigger rate is plotted in Fig.~\ref{fig:SAA-1}. The trigger rate shows a sharp increase in the SAA region. An appropriate threshold of this trigger rate is chosen to determine when the satellite enters and exits the SAA, and the boundary of the SAA is defined as the corresponding positions of the satellite. Events detected between the time when the satellite enters and exits the SAA are tagged.

This method for identifying the SAA region is directly based on the particle flux in space. With such a way we can dynamically discard the ``bad" data in the SAA region and keep as many good events as possible, thereby improving the efficiency for analysis \cite{JiangW2017}. Meanwhile, based on this subtle method, the variations of SAA with solar and geomagnetic activities can also be studied.

\section{Live time}

A proper estimation of the live time is very important for the cosmic ray and $\gamma$-ray flux measurements. For the DAMPE detector, the live time is mainly governed by three factors. The first one is the SAA region. The definition of the SAA has been introduced in Section 8. DAMPE travels through the SAA about 7 times per day and loses about $5\%$ of the operation time. The second factor is the instrumental dead-time. For the nominal science operation, the instrumental dead-time is mainly due to the trigger generation, the signal readout, the data transform, and the electronics recovery, which yields a total dead-time of 3.0725 ms per event. Considering the trigger rate outside the SAA region, this factor gives a live time loss of about $18.44\%$ of the operation time. The third one is due to the on-orbit calibration. DAMPE performs on-orbit calibration of pedestal and electronics linearity periodically, which in total leads to a live time loss of about $1.56\%$. Thanks to the reliability of the data acquisition and transmission system, the loss of live time due to the data loss is negligible.

\section{Payload internal and spacecraft alignment}

\subsection{The alignment of PSD}

The alignment of the PSD is important for improving the charge resolution and the $\gamma$-ray identification efficiency. After the proper PSD alignment, the trajectories of the incident particles in PSD bars are better tracked and the deposited energies in PSD are more accurately reconstructed.
We select the proton MIP events to study the effect of misalighment. Generally speaking there are 6 degrees of freedom for each PSD bar, i.e., the shifts in the $XY$ plane (one longitudinal and one transverse) and in the $Z$ direction, and the rotations in $XY$, $YZ$, $XZ$ planes.
We divide each PSD bar (800 mm) uniformly into 11 segments. By setting more restrictive geometry conditions, the proton MIP events that passing through the whole PSD bar are selected. It is found that the charge distribution of selected events agree well with the MC results where the ideal geometry model is employed, implying that the rotations in $XZ$ and $YZ$ planes are negligible. In addition, according to the event reconstruction procedure the particles that pass through the edges of the PSD detector are rejected, the longitudinal shift of PSD bars can also be ignored.
Therefore, we only focus on the rest 3 degrees of freedom for each PSD bar: the rotation in $XY$ plane, shift in $Z$ axis direction, and transverse shift in $XY$ plane. 


\begin{figure}
\includegraphics[scale=0.22]{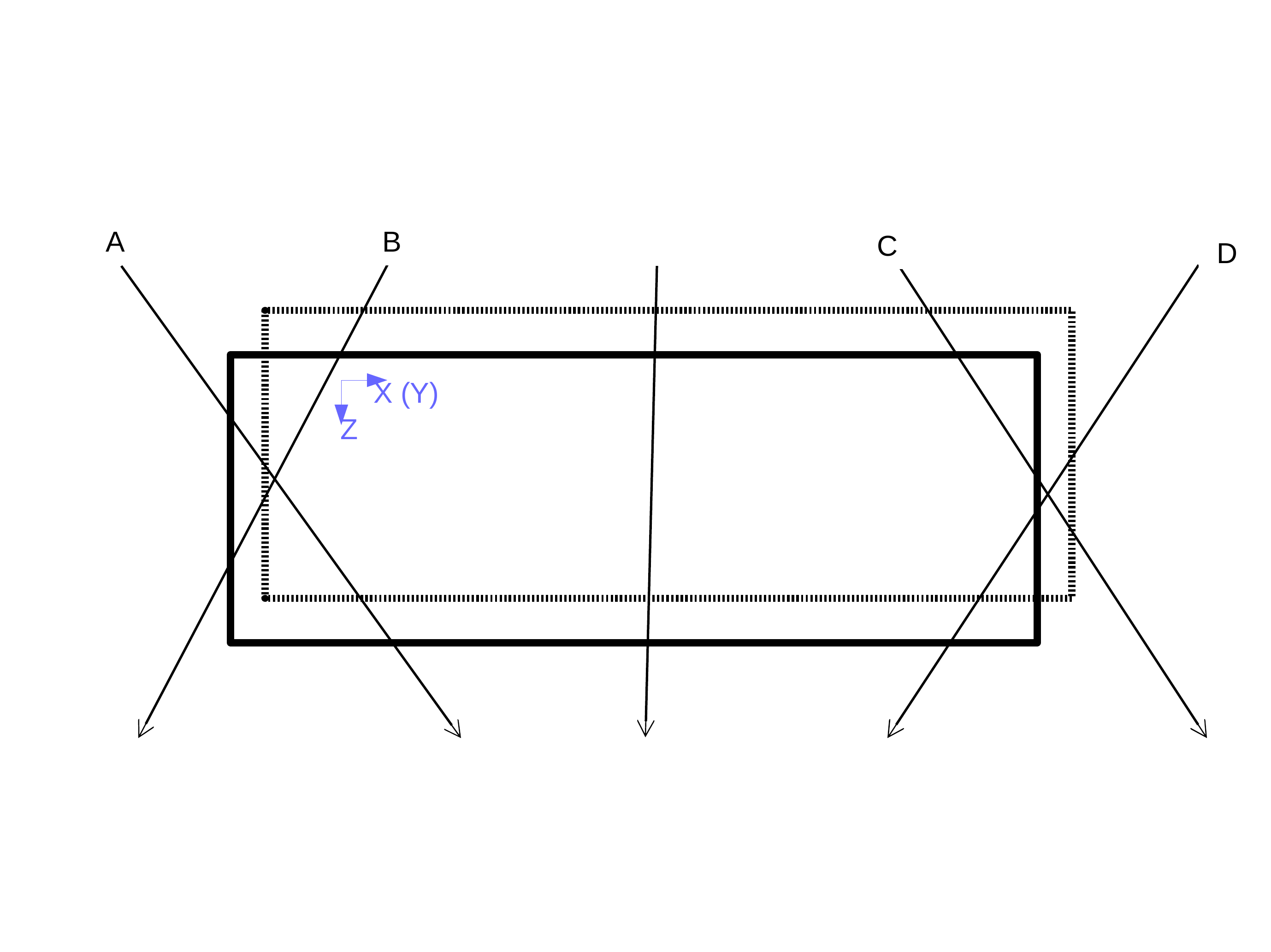}
\includegraphics[scale=0.30]{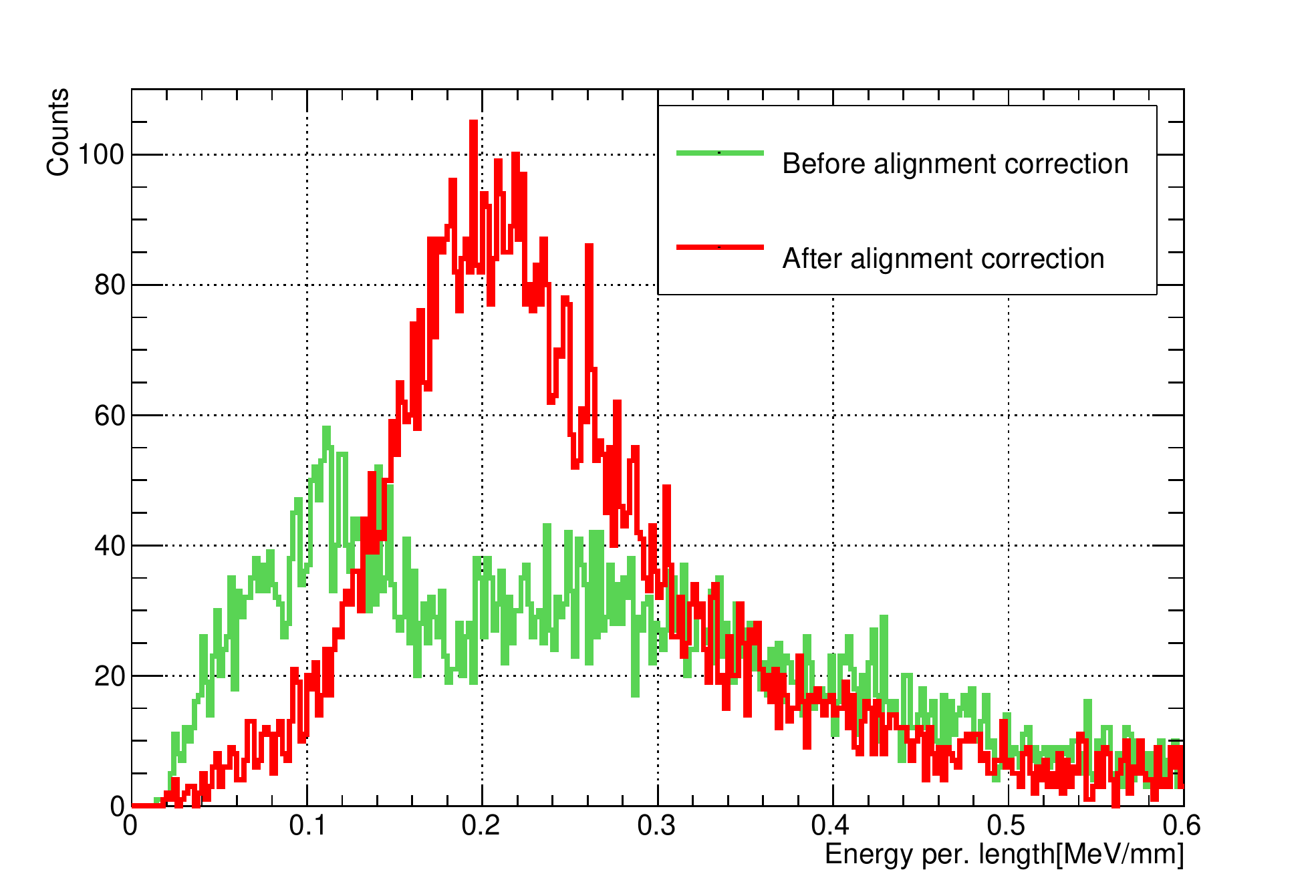}
\caption{The left panel shows a slight misalignment of a PSD bar from the designed position (dash-line rectangle) to the real position (solid line rectangle) distorts the path length measurements of the tracks (A, B, C and D) that pass through the PSD bar from side. A comparison of the deposit energy per unit length of the A-, B-, C-, and D-type tracks before (green histogram) and after (red histogram) alignment of one particular segment is shown in the right panel.} \label{fig:PSD-alig}
\end{figure}

As shown in the left panel of Fig.~\ref{fig:PSD-alig}, for each proton MIP event like $A$, $B$, $C$ and $D$, we can calculate the path length and the energy deposition per millimeter as a function of the three misalignment parameters. Through setting the energy deposition per millimeter to the standard value of 0.2 MeV/mm, we can derive the misalignment parameters using an iteration algorithm. The results are breifly summarized in Table \ref{tab-PSD-alig}.

The influence of the alignment correction to the energy measurement is illustrated in the right panel of Fig.~\ref{fig:PSD-alig}. It is clear that this correction is important for particles that pass through the corners of PSD bars. As expected, after applying the PSD alignment, the charge resolution for all kinds of cosmic ray nuclei is improved considerably. We refer the readers to Refs. \cite{MaPX2018,DongTK2018} for more details about the PSD alignment procedure and the charge measurement results.



\begin{tiny}
\begin{table}[!htb]
\begin{center}
\caption{The alignment parameters of PSD.}\label{tab-PSD-alig}
\begin{tabular}{llllll}
\hline
& Vertical shift range (mm) & Horizontal shift range (mm) & Rotation in XY plane range (rad)  \\
\hline
 The first layer    & $-2.72\sim -1.48$ &	$-0.18\sim 0.56$ &	$0.0012\sim0.0021$     \\
 The second layer	& $-0.57\sim -1.21$ &	$0.21\sim 0.88$  &	$-0.0018\sim-0.0007$       \\
\hline
\end{tabular}
\end{center}
\end{table}
\end{tiny}

\subsection{The alignment of STK}
\label{sec:stk_alignment}

The construction precision of the mechanical assembly of the STK is O(100) $\mu \mathrm{m}$. Given the expected position resolution of silicon sensors better than 70 $\mu \mathrm{m}$, a precise alignment of the instrument is required. Aligned positions of hits in the STK are expressed though their ideal positions as follows
\begin{align}
x_{a} &= x + \Delta_x - y \cdot \theta_{z} \\
y_{a} &= y + \Delta_y + x \cdot \theta_{z} \\
z_{a} &= z + \Delta_z - x \cdot \theta_{y} + y \cdot \theta_{x}
\end{align}
where $\Delta_x/\Delta_y$, $\Delta_z$, $\theta_{x}$, $\theta_{y}$, $\theta_{z}$ are the alignment parameters for shifts and rotations of each silicon sensor, respectively. There are a total of 3840 alignment parameters, for 768 silicon sensors.

Alignment of the STK is done is through the minimization of the global $\chi^2$ of the reconstructed tracks in the alignment sample. Tracks used in the alignment are required to have hits in all six $X$ and $Y$ planes of the STK. Each track in the alignment sample is re-fitted with a straight line, and a corresponding $\chi^2$ is evaluated and added to a global $\chi^2$.   Tracks which yield $\chi^2$ higher than a certain (conservatively high) threshold are removed from the sample.
A custom implementation of the gradient optimization algorithm is used for the alignment, as described in ~\cite{stk_alignment_paper}.
Alignment of the STK is performed every two weeks and most up-to-date alignment constants are used in the data processing for the track reconstruction. A sample of around $10^6$ tracks is used at each update of the alignment, which corresponds to about one and a half day of the orbit data.

\begin{figure}[h]
\includegraphics[width=0.8\textwidth]{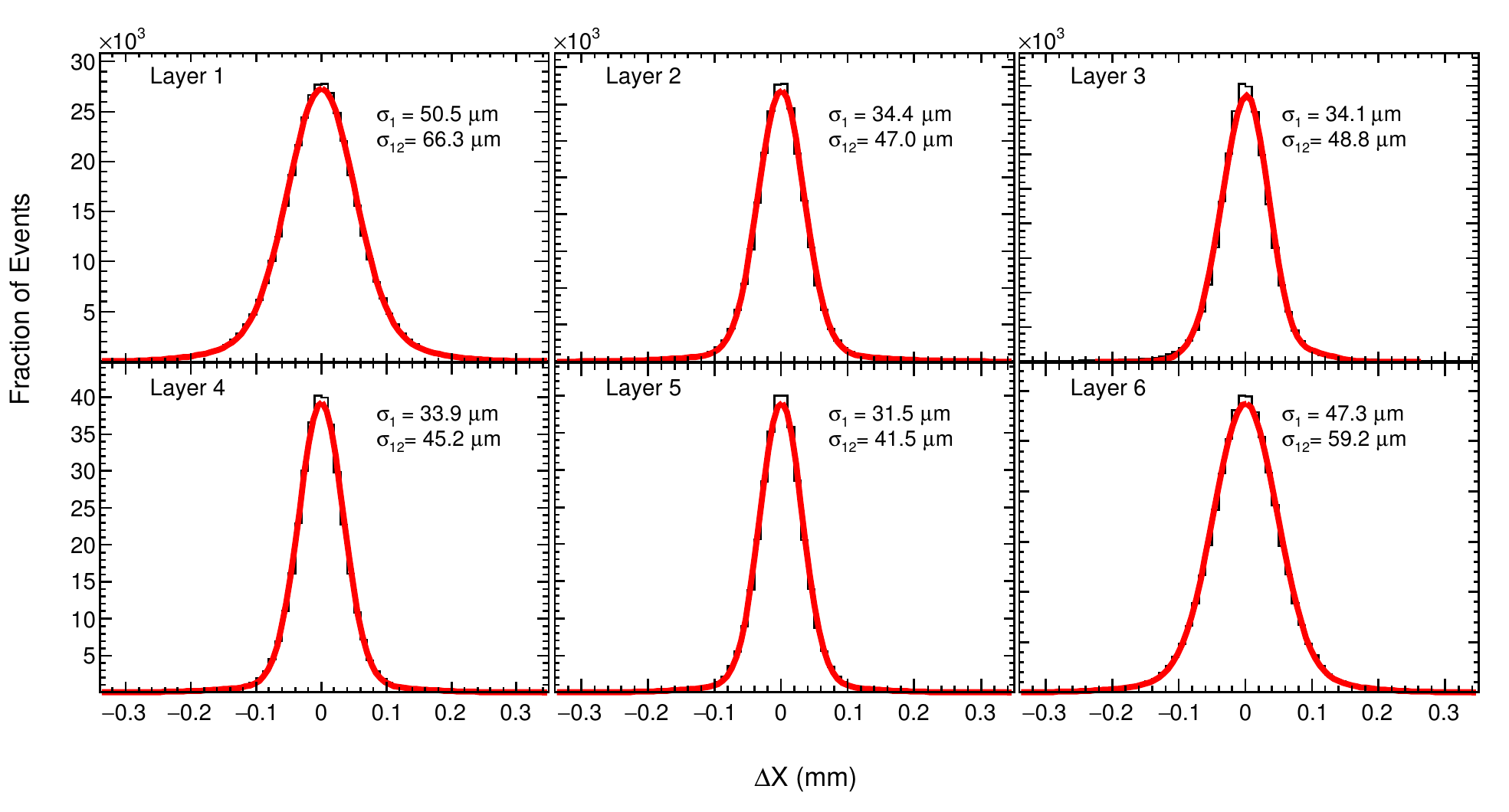}
\includegraphics[width=0.8\textwidth]{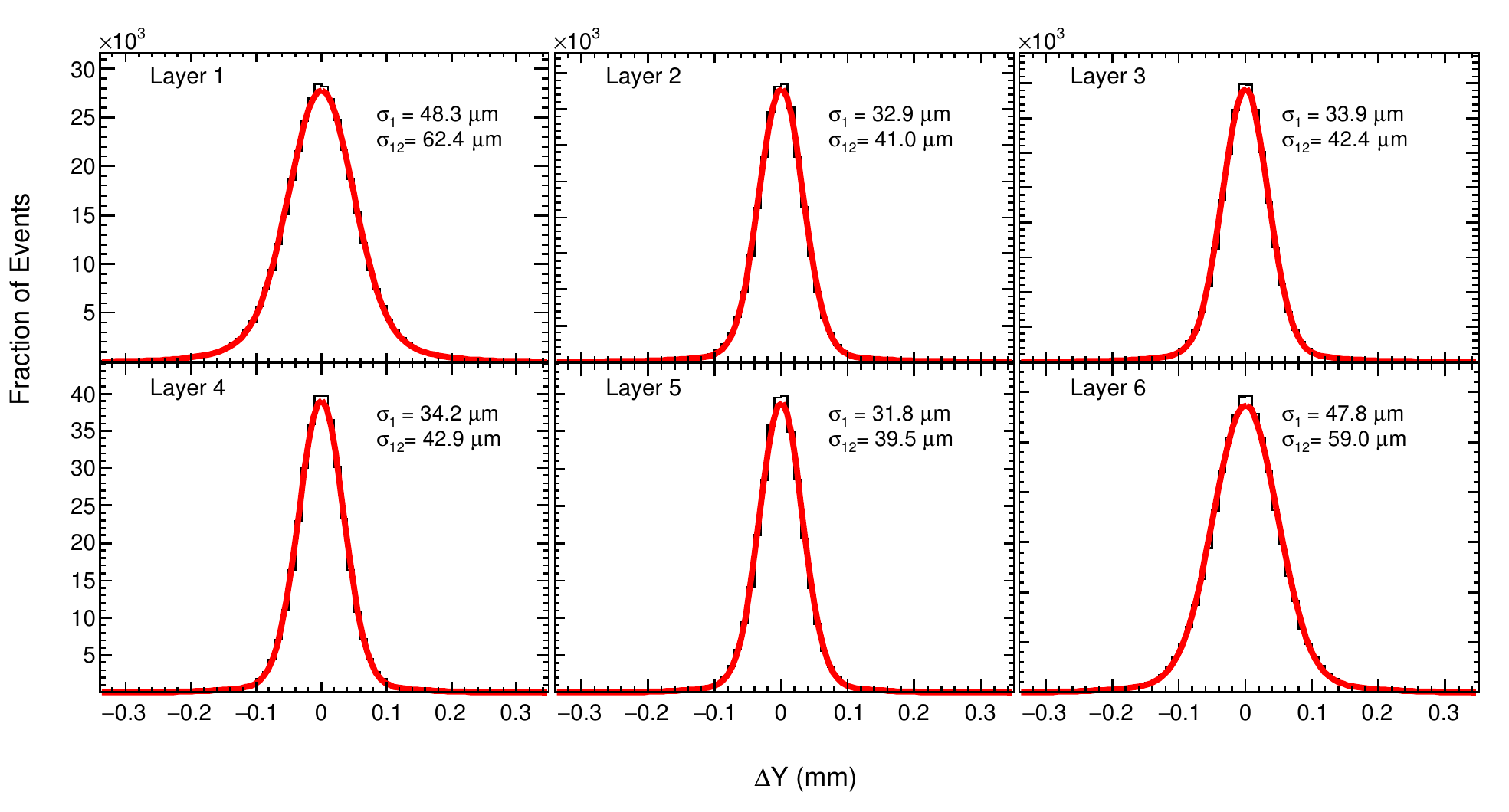}
\caption{ Residual distribution of projected minus measured hit position in a track after the alignment. The tracks with all incidence angles are selected using the tight criterion. Projected position is obtained from a 5-point linear fit, excluding the plane itself where residual is calculated. Statistics in the plots corresponds to five subsequent days of orbit data. Distributions are fitted by the double-Gaussian function~\cite{Alpat:2010zz}.}
\label{fig:stk_alignment_residue}
\end{figure}

The position resolution of the STK is used as a measure of the performance of the alignment. To estimate the position resolution, a sub-set of tracks is used with the tight selection criteria, as described below. Then the residuals are evaluated defined as difference between projected and measured coordinate of a track in the $i$-th plane, where the projection is obtained by using the remaining five points of a track not including the point in the $i$-th  plane.
The tight selection is defined as follows: the track residuals in each point of the 5-point fit must be lower than a certain threshold,  O(10) $\mu \mathrm{m}$. The latter helps to eliminate events which suffer most from multiple scattering. The resulting residuals are shown in the histograms for different STK layers (Fig.~\ref{fig:stk_alignment_residue}).

\begin{figure}[]
\includegraphics[width=0.8\textwidth]{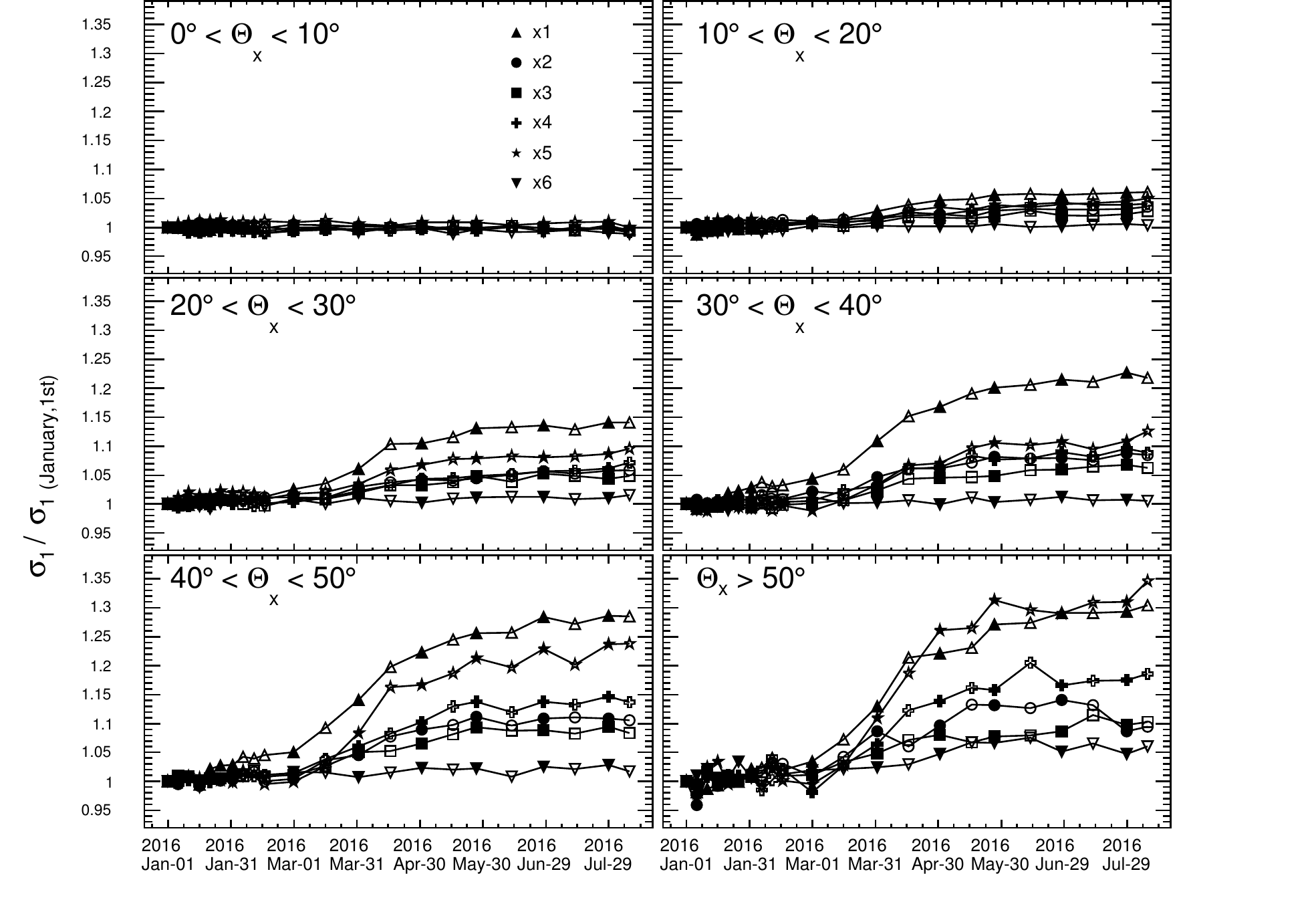}
\includegraphics[width=0.8\textwidth]{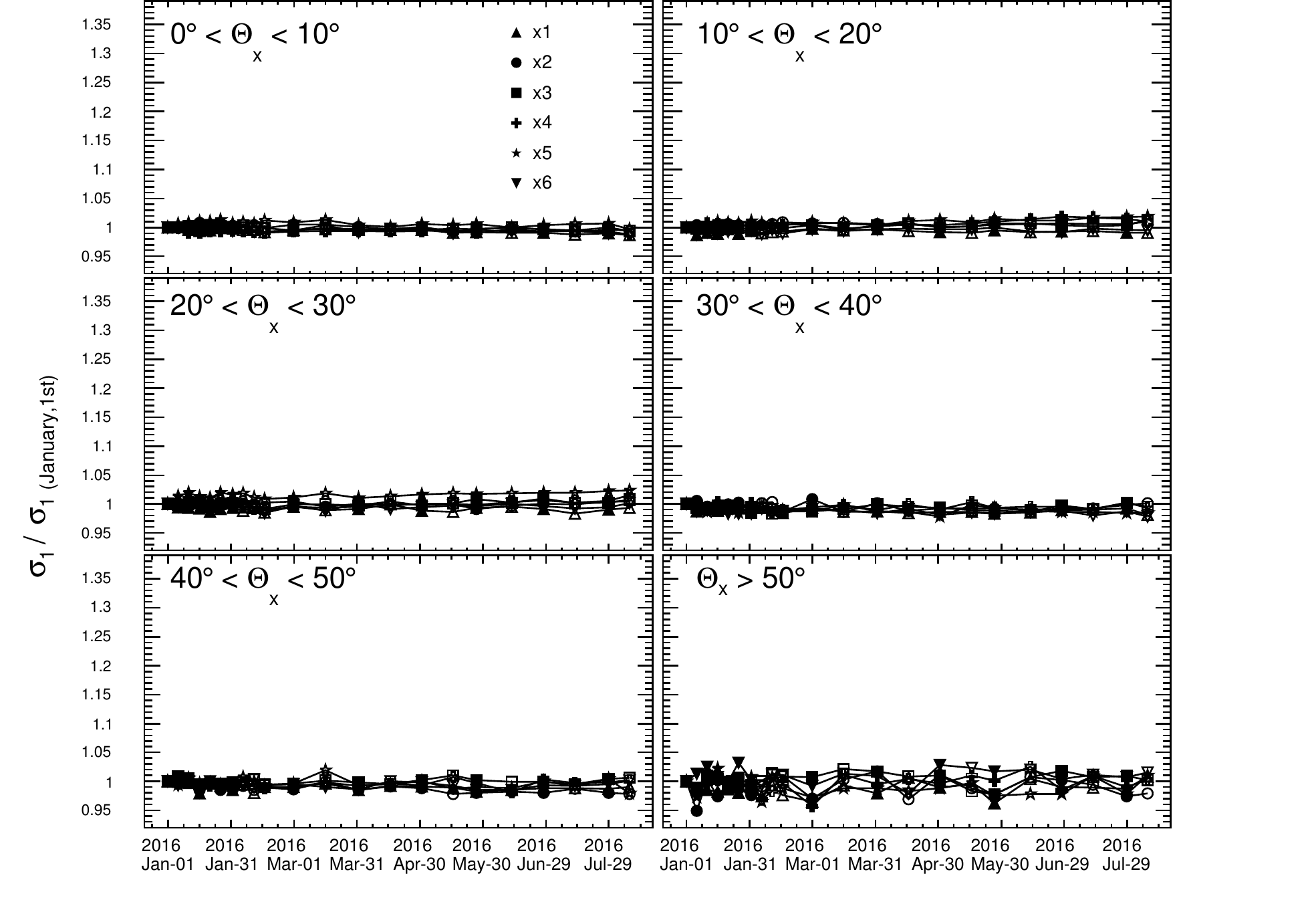}
\caption{The RMS values of the major Gaussian in the fit of track-hit residuals shown as functions of time for different STK planes at different track inclinations. The RMS are normalized to their corresponding values determined on \stkjanuaryfirst. The fixed (top) and time-dependent (bottom) alignments are shown.}
\label{fig:stk_alignment_stability}
\end{figure}

To assess the stability of the alignment, we examine the position resolution at different time periods on orbit. As a measure of the stability of the alignment we use the RMS of the major Gaussian in the fit of the residuals. The time behavior of this parameter is shown in Fig.~\ref{fig:stk_alignment_stability} (top and bottom) for the fixed and time-dependent alignment, respectively. In the former case, one single alignment is performed in January 2016 and used for the whole orbit data set. In the latter case, alignment parameters are updated once per two weeks, and the values of  the closest-in-time alignment parameters are used for each track in the  sample. We have also checked the effect of performing the alignment once per two days as well as once per week  and found it to have no significant improvement with respect to the nominal procedure, in which alignment is done every two weeks.

\subsection{Boresight alignment of DAMPE}

The direction of each detected particle is reconstructed with respect to the reference system of the DAMPE payload.
To achieve the celestial coordinate of a particle, the transformation from the payload coordinate system to the celestial coordinate system is required. The transformation is determined by the orbit parameters and the celestial orientation of the satellite which are provided by the Navigation system and star-tracker, respectively. Small deviations from real pointing may be introduced to the transformed celestial coordinate, due to thermal variations, acoustic vibrations, 0g fluctuations, and the uncertainties in the orbit parameters and star-tracker pointing. This mis-match will not only cause a systematic shift between the mean observed position and the real position of a point source, but also lead to a distorted point spread function (PSF) profile. In this section, we use the gamma-ray data centered around several bright point sources to measure and correct the angular deviation from the real celestial coordinate, the so called  ``boresight alignment'' of the DAMPE payload.

The boresight alignment of the payload can be expressed using a series rotation about the x, y and z axes of the satellite, where  z axis is opposite zenith, y axis is along the normal direction of the orbital plane, and x axis is determined by z axis and y axis accordingly. Specifically, the rotation of a vector $\hat{\bm r}$ in the payload coordinate system can be absolutely described by three angles $(\theta_{\rm x}, \theta_{\rm y}, \theta_{\rm z})$, which can be written as
\begin{equation}
\hat{\bm r}'({\bm \theta}) \equiv \hat{\bm r}'(\theta_{\rm x}, \theta_{\rm y}, \theta_{\rm z}) = R_{\rm x}(\theta_{\rm x}) R_{\rm y}(\theta_{\rm y}) R_{\rm z}(\theta_{\rm z}) \hat{\bm r},
\end{equation}
where $R_{\rm x}$, $R_{\rm y}$ and $R_{\rm z}$ are the rotation matrices along x, y and z axes respectively, defined as
\begin{equation}
R_{\rm x}(\theta_{\rm x}) = \left[
\begin{matrix}
1&0&0&\\
0&\cos \theta_{\rm x}&\sin \theta_{\rm x}&\\
0&-\sin \theta_{\rm x}&\cos \theta_{\rm x}&
\end{matrix}
\right],
\end{equation}
\begin{equation}
R_{\rm y}(\theta_{\rm y}) = \left[
\begin{matrix}
\cos \theta_{\rm y}&0&-\sin \theta_{\rm y}&\\
0&1&0&\\
\sin \theta_{\rm y}&0&\cos \theta_{\rm y}&
\end{matrix}
\right],
\end{equation}
\begin{equation}
R_{\rm z}(\theta_{\rm z}) = \left[
\begin{matrix}
\cos \theta_{\rm z}&\sin \theta_{\rm z}&0&\\
-\sin \theta_{\rm z}&\cos \theta_{\rm z}&0&\\
1&0&0&
\end{matrix}
\right].
\end{equation}

Since the rotation will transform the photon direction in the payload coordinate, it will result in a different direction in celestial coordinate. If $R_{\rm sky}(t)$ is the rotation matrix from the payload coordinate to the celestial coordinate at a certain time $t$, the correction of a vector $\hat{\bm p}$ in celestial coordinates due to boresight alignment can be expressed as
\begin{equation}
\label{celestial_correction}
\hat{\bm p}'({\bm \theta}) = R_{\rm sky}(t) R_{\rm bor}({\bm \theta}) R_{\rm sky}^{-1}(t) \hat{\bm p},
\end{equation}
where $R_{\rm bor}({\bm \theta}) \equiv R_{\rm x}(\theta_{\rm x}) R_{\rm y}(\theta_{\rm y}) R_{\rm z}(\theta_{\rm z})$.

The boresight alignment is determined via a likelihood maximization for some bright gamma-ray point sources. Particularly, we select gamma-rays from Vela pulsar which is the brightest point gamma-ray source in the sky to determine these three rotation angles. We use the DAMPE data from January 1 2016 to January 1 2017 for the analysis. We select the photons within 4$^\circ$ of the real position of Vela and restrict the energies to be between 3 GeV and 100 GeV. The gamma-rays in such a region of interest (ROI) can be modeled by a point source and a uniform background template. The background spectrum is further modelled by a power-law.
The spectral and spatial parameters of the point source is imported from the third Fermi-LAT catalog of high-energy sources \cite{3FHL}.
Therefore, the likelihood is determined by the three alignment angles and the spectral parameters of the background, in the form of
\begin{equation}
{\rm ln}\,\mathcal{L}( \lambda, {\bm \theta}) =
- \iiint_{\rm ROI} \sum_{\rm j=1}^{n_{\rm sources}} r_{j}\left (E', \hat{\bm p}; t', \vec{\lambda}_{j}
 \right ) + \sum_{\rm i=1}^{n_{\rm events}} {\rm ln} \left (
 \sum_{\rm j=1}^{n_{\rm sources}} r_{j} \left (
 E_{i}', \hat{\bm p}_{i}' ( {\bm \theta}); t', \vec{\lambda}_{j}
 \right ) \right ),
\end{equation}
where $r_{j}$ is the expected contribution from the $j$-th source to the photon with constructed energy $E'$ and direction $\hat{\bm p}$ in celestial coordinate system, $t'$ is exposure time,  $\vec{\lambda_{j}}$ is the parameter space of the $j$-th source, and $\hat{\bm p}'_{\rm i}$ is the direction of i-th photon after boresight alignment. We use the {\tt MINUIT} algorithm\footnote{https://github.com/iminuit/iminuit} to perform the optimization \cite{minuit}, and yield the boresight alignment constants to be
\begin{equation}
\theta_{\rm x} = 0.126^\circ \pm 0.013^\circ,\
\theta_{\rm y} = 0.021^\circ \pm 0.009^\circ,
\theta_{\rm z} = -0.143^\circ \pm 0.012^\circ.
\end{equation}

The same procedure is also performed using the gamma-rays from Geminga and Crab pulsars, and consistent results are obtained. More statistics are still needed  to improve the precision.

\section{Conclusion }

The methodology of the on-orbit calibration adopted for DAMPE has been introduced in this paper, including the pedestals, the zero-suppression thresholds, the trigger thresholds, the calculation of detector gains, the MIP responses, the light attenuation, the determination of the SAA region, the measurement of the live time, and the detector alignments.

The calibration results form a solid foundation for data analysis. All the on-orbit calibration results are consistent with the ground test results, indicating that DAMPE is functional in space as expected. Furthermore, all the calibration aspects, including the pedestals, the dynode ratios, and the energy gains etc, are very stable after the temperature correction. Based on the experience of calibration in the first 15 months and the good performance of all sub-detectors, DAMPE produces high quality science data (e.g., \cite{DAMPE2017,LeiSJ2017,LiangYF2017,YueC2017,Gallo2017}) and is expected to operate stably during the next few years.


\section*{Acknowledgement}
{ We thank the referee for helpful suggestions.}
The DAMPE mission was sponsored by the strategic priority science and technology projects in space science of the Chinese Academy of Sciences. In China the data analysis is supported in part by National Key R\&D Program of China (No. 2016YFA0400200), the National Basic Research Program (No. 2013CB837000), NSFC under grants No. 11525313 (i.e., Funds for Distinguished Young Scholars), No. 11622327 and No.11722328 (i.e., Funds for Excellent Young Scholars), No. 11273070, No. 11303096,  No. 11303105, No. 11303106, No. 11303107, No.11673021, No. 11673047, No.11673075, No.11705197, No.11773075, No.11773085, No.11773086, U1531126, U1631111, U1738121, U1738123, U1738127, U1738129, U1738133, U1738135, U1738136, U1738137, U1738138, U1738139, U1738205, U1738206, U1738207, U1738208, U1738210 and the 100 Talents program of Chinese Academy of Sciences. In Europe the activities and the data analysis are supported by the Swiss National Science Science Foundation, the Italian National Institute for Nuclear Physics (INFN), and the Italian University and Research Ministry (MIUR).



\section*{References}

\end{document}